\documentclass[fleqn,usenatbib]{mnras}

\usepackage{newtxtext,newtxmath}
\usepackage[T1]{fontenc}

\DeclareRobustCommand{\VAN}[3]{#2}
\let\VANthebibliography\thebibliography
\def\thebibliography{\DeclareRobustCommand{\VAN}[3]{##3}\VANthebibliography}

\usepackage{graphicx}	% Including figure files
\usepackage{amsmath}	% Advanced maths commands
\newcommand{\iso}[2]{\hbox{${}^{#1}{\rm #2}$}}
\newcommand{\Msun}{\ensuremath{{M}_{\odot}}}
\newcommand{\Lsun}{\ensuremath{{\rm L_{\odot}}}}
\newcommand{\Rsun}{\ensuremath{{\rm R_{\odot}}}}

\usepackage{cancel}

\title[Most metal-rich AGB stars]{The most metal-rich asymptotic giant branch stars}

\author[Karakas, Cinquegrana, \& Joyce]{
Amanda I. Karakas,$^{1,2}$\thanks{E-mail: amanda.karakas@monash.edu}, Giulia Cinquegrana,$^{1,2}$ and Meridith Joyce,$^{2,3,4}$
\\
$^{1}$School of Physics \& Astronomy, Monash University, Clayton VIC 3800, Australia\\
$^{2}$ARC Centre of Excellence for All Sky Astrophysics in 3 Dimensions (ASTRO 3D)\\
$^{3}$Research School of Astronomy \& Astrophysics, Mount Stromlo Observatory, the Australian National University, Canberra ACT 2611, Australia\\
$^{4}$Space Telescope Science Institute, 3700 San Martin Drive, Baltimore, MD 21218, USA\\
}

\date{Accepted XXX. Received YYY; in original form ZZZ}

\pubyear{2020}

\begin{document}
\label{firstpage}
\pagerange{\pageref{firstpage}--\pageref{lastpage}}
\maketitle

\begin{abstract}
We present new stellar evolutionary sequences of very metal-rich stars evolved with the Monash Stellar Structure code and with MESA.
The Monash models include masses of $1-8\Msun$ with metallicities $Z=0.04$ to $Z=0.1$ and are evolved from the main sequence to
the thermally-pulsing asymptotic giant branch (AGB). These are the first $Z=0.1$ AGB models in the literature. The MESA models include
intermediate-mass models with $Z=0.06$ to $Z=0.09$ evolved to the onset of the thermally-pulsing phase. Third dredge-up only
occurs in intermediate-mass models $Z \le 0.08$. Hot bottom burning (HBB) shows a weaker
dependence on metallicity, with the minimum mass increasing from 4.5$\Msun$ for $Z=0.014$ to $\approx 5.5 \Msun$ for Z = 0.04,
$ 6 \Msun $ for $ 0.05 \le Z \le 0.07$ and above 6.5$\Msun$ for $Z\ge 0.08$. The behaviour of the $Z=0.1$ models is unusual;
most do not experience He-shell instabilities owing to rapid mass-loss on the early part of the AGB.
Turning off mass-loss produces He-shell instabilities, however thermal pulses are weak and result in no third dredge-up.
The minimum mass for carbon ignition is reduced from 8$\Msun$ for $Z=0.04$ to 7$\Msun$ for $Z=0.1$, which implies a reduction in the
minimum mass for core-collapse supernovae. MESA models of similarly high metallicity ($Z=0.06 - 0.09$) show the same lowering of the minimum
mass for carbon ignition: carbon burning is detected in a $6\Msun$ model at the highest metallicity ($Z=0.09$) and in all $7\Msun$ models with $Z \ge 0.06$. This demonstrates robustness of the lowered carbon burning threshold across codes.
\end{abstract}


%%%%%%%%%%%%%%%%%%%%%%%%%%%%%%%%%%%%%%%%%%%%%%%%%%

%%%%%%%%%%%%%%%%% BODY OF PAPER %%%%%%%%%%%%%%%%%%

\section{Introduction}

Low and intermediate-mass stars evolve through core hydrogen and helium burning before ascending the asymptotic giant branch phase (AGB). It is during the AGB that the richest nucleosynthesis occurs, driven by mixing episodes between the core and envelope. Prior to the thermally-pulsing AGB (TP-AGB), first and second dredge-up may also occur, which change the surface composition by bringing the products of hydrogen burning to the surface.  For reviews of the evolution of low and intermediate-mass stars we refer to \citet{iben83}, \citet{busso99}, \citet{herwig05}, and \citet{karakas14dawes} and references therein.

Stellar yields from thermally-pulsing AGB stars are shaped by the repeated action of the third dredge-up (TDU), which mixes the products of partial He-burning to the stellar envelope, and hot bottom burning (HBB). TDU can cause the surface composition to become carbon rich, where the number of carbon atoms exceeds the number of oxygen atoms, i.e., C/O $\ge 1$. The minimum stellar mass for TDU depends on initial parameters such as mass and metallicity, and is $\approx 1.5\Msun$ for solar metallicity \citep[e.g.,][]{groen95,marigo20}. 
There is evidence for TDU occurring in low-mass Galactic AGB stars of $\approx 1\Msun$, which is a challenge to model if we find that these stars are around solar metallicity \citep{shetye19}. 

Theoretical models predict that TDU has a strong metallicity dependence \citep{boothroyd88c,straniero97,karakas02}.
Low-metallicity stellar models show more efficient TDU as well as a lower minimum mass, while above solar metallicity we predict the reverse, that the minimum mass increases and the efficiency at a given mass is reduced \citep{karakas14b,weiss09,ventura20}. What happens when the metallicity is greater than about twice the solar value? There are currently too few AGB models in the literature to answer this question. 

HBB occurs when the base of the convective envelope becomes hot enough to sustain proton capture nucleosynthesis \citep{bloecker91,boothroyd92,lattanzio92,ventura13}. The efficiency of HBB depends on the temperature and density gradients near the base of the envelope, and on the peak temperature in that region. These quantities depend on the total stellar mass, core mass, and metallicity as well as on the convective model adopted in the stellar evolution code \citep[e.g.,][]{ventura05a}.

The minimum initial mass for HBB is approximately 5$\Msun$ for solar metallicity \citep{colibri,karakas14b}. However different codes predict different minimums. For models around solar metallicity, the minimum varies from 3.5$\Msun$ \citep[e.g.,][]{ventura18} to 6$\Msun$ \citep{weiss09} or even no HBB at all \citep{straniero14}. All codes predict the same qualitative trend with metallicity in that the minimum mass decreases as a function of decreasing metallicity, and at a given mass the maximum temperature reached at the base of the envelope increases. The minimum mass for HBB for very metal-rich AGB stars is uncertain, as there are few models in the literature.

In \citet{karakas14b} we present grids of low and intermediate-mass mass stars between $1-8\Msun$ of solar metallicity, and a factor of two above and below solar. Solar metallicity in that paper and here is defined as $Z = 0.014$ \citep{asplund09}, so the full metallicity range considered is $Z = 0.007, 0.014$ and $0.03$. In a follow-up paper, \citet{karakas16} provide the first grid of super-solar AGB yields, including the $s$-process, for AGB stars of $Z=0.03$.    Interestingly, the mass range of carbon stars narrowed from 1.5 to 4.5$\Msun$ to 3 to 4$\Msun$ when the metallicity was increased by a factor of two above solar, and the minimum mass for HBB increased to 5$\Msun$ \citep[also see models by][]{marigo13}.

There is a metallicity ceiling to carbon star production \citep{boyer13, boyer19,marigo17}, which limits the amount of carbon that AGB stars can make as a function of increasing metallicity. Here, the metallicity ceiling is a result of two reasons: 1) the increasing amount of oxygen with increasing $Z$ means it is more difficult to dredge-up enough carbon to reach  C/O $\ge 1$, and 2) the predicted TDU efficiency decreases from solar to $Z=0.03$ \citep{karakas14b}. What happens in very metal-rich AGB stars above twice solar? Could the TDU simply not occur in low-mass stars below about 4$\Msun$?

Thinking more generally, is there an upper metallicity limit above which AGB nucleosynthesis does not enrich the stellar surface? Here we include the nucleosynthesis contribution from HBB, which occurs in intermediate-mass AGB models of $\approx 4-8\Msun$ at solar metallicities.  Or indeed, perhaps a metal-rich threshold for low and intermediate-mass stars to enter the AGB phase itself? If so, this means that above some metallicity threshold, AGB stars no longer contribute to the chemical evolution of carbon and $s$-process elements through single stellar evolution. In this study we aim to answer these questions by considering models of the most metal-rich AGB stars. 

These questions include: (1) how do very metal-rich models of low and intermediate-mass stars compare to their lower metallicity counterparts? (2) Are there significant changes to stellar lifetime, to the depth of first and second dredge-up, to the extended radii on the giant branches? On the AGB phase we need to consider the strength of thermal pulses, the temperatures in the burning shells, and on the efficiency of mixing episodes. How do the minimum masses for TDU and HBB depend on mass and metallicity, as the metallicity increases well above solar? 

These results carry important implications for the composition of dust production in galaxies \citep[e.g.,][]{lugaro20,ventura20}, and for the chemical evolution of galaxies \citep{romano10,kobayashi20}.  In particular, the inner regions of spiral galaxies such as our Milky Way Galaxy and M~31, along with giant ellipticals may harbour very metal-rich stellar populations \citep{worthey94,ness16,do18,thorsbro20}. Stellar evolution fitting of metal-rich star clusters 
\citep[e.g., NGC 6791, NGC 6253 and NGC 6583;][]{carraro06,netopil16}, all of which have [Fe/H] $\approx +0.4$, requires models of stars at these metallicities. Very metal-rich stellar models are also required to calculate the survival rates of planets around metal-rich host stars \citep{villaver07}, where gas giant planets in particular are found in greater numbers around stars with increasing metallicity \citep[e.g.,][]{fischer05}.

Here we present stellar evolutionary models of low and intermediate-mass stars between 1--8$\Msun$ with $Z = 0.04$ to $Z=0.1$ including the full thermally-pulsing AGB phase.  This is the first set of AGB models of $Z=0.1$ presented in the literature. While we will focus on the evolution during the AGB phase, we examine the behaviour of the first and second dredge-up at these metallicities, stellar lifetimes, and the minimum mass for core carbon ignition. The minimum mass for core carbon burning ultimately determines which stars become white dwarfs or go onto explode as core collapse supernova \citep{bono00,ibeling13,doherty17}. 

The AGB models are calculated with the Monash Stellar Structure code \citep{lattanzio86,frost96,karakas14b} however the implications for carbon ignition, which shift the boundaries for core-collapse supernovae, require careful checking. As noted by \citet{doherty14b}, the difficult phase of carbon ignition in the degenerate core of an intermediate-mass AGB star is best done using a diffusive mixing scheme \citep[see also][]{siess06,gilpons13}. The version of the Monash Stellar Structure code that we use does not include diffusive mixing and instead uses instantaneous mixing in all convective regions \citep[e.g., as discussed in][]{karakas14b}. Therefore, it is important to check that our finding of the minimum mass for C-ignition shifting downwards by $\approx 1\Msun$ at the highest $Z$ is not a result of input physics or numerical choices. We note that previous studies of metal-rich models that explored this boundary did not comment on this feature \citep[e.g.,][]{mowlavi98a} or did not find the turnover in minimum C-ignition mass with increasing $Z$ \citep{ibeling13}.
For these reason we compute a supplementary grid of stellar tracks with MESA covering metallicities between $Z=0.06$ to 0.09 and masses between 4$\Msun$ to 9$\Msun$. The advantage of MESA is that it includes options for diffusive mixing and convective boundary mixing, and has been successfully used in the 8-10$\Msun$ range by e.g,. \citet{jones13,jones14}.

We begin in \S~\ref{sec:literature} with a review of the metal-rich models available in the literature.  Section~\ref{sec:models} then introduces the new stellar evolutionary sequences, while our results are summarized in \S~\ref{sec:results}.  We finish with a discussion in \S~\ref{sec:discuss}, and our conclusions in \S~\ref{sec:conclusion}.

\section{Metal-rich models in the literature} \label{sec:literature}

\citet{meynet06} review the evolution of very metal-rich stars, based partly on models from \citet{mowlavi98a} and \citet{mowlavi98b}.  While the \citet{meynet06} review gives a good overview of the evolution and properties of very metal-rich stars including up to $Z=0.1$, the discussion concentrates on massive star evolution. There are a number of studies that focus on low and intermediate-mass stellar evolution between $Z=0.04$ to $Z=0.1$ but these do not evolve through the thermally-pulsing AGB phase \citep{fagotto94a,fagotto94b,mowlavi98a,mowlavi98b,bono00,salasnich00,claret07,valcarce13}.

Models of very metal-rich thermally-pulsing AGB stars are fewer in number. \citet{siess07} modelled the evolution of stars with masses between 9-13$\Msun$, which experience off-centre carbon ignition and develop O-Ne cores before evolving through the super-AGB phase. The most metal-rich models were $Z=0.04$, which is $2 Z_{\odot}$ where he set $Z_{\odot} = 0.02$.   Models of CO-core AGB stars include those by \citet{bertelli08} who used a synthetic treatment of the thermally-pulsing AGB phase of evolution; \citet{weiss09} who model AGB stars up to $Z=0.04$; \citet{colibri} model AGB stars with metallicities up to $Z=0.05$ and \citet{marigo17} present results for metallicities up to [Fe/H] = $+0.7$; the {\tt MIST} models by \citet{choi16} up to [Fe/H] = $+0.5$; and the $Z = 0.03$ and $Z = 0.04$ AGB models by \citet{ventura20}. Similar to \citet{siess07}, \citet{weiss09} and \citet{ventura20} have a maximum $2 Z_{\odot}$, where the abundances used in the calculations are based on \citet[][that is, $Z=0.02$]{grevesse98}.  

\citet{marigo17} adopt $Z_{\odot} = 0.01524$ which means that their most metal-rich models have $Z\approx 0.08$. The {\tt MIST} evolutionary models use the same solar metallicity as is adopted here, which means that their most metal-rich models have $Z\approx 0.05$ (assuming a solar scaling of elements). There are currently no AGB models in the literature with $Z=0.1$.

\section{The Stellar models} \label{sec:models}

Here we describe the models calculated and the codes used to calculate them.

\subsection{The Monash Stellar Models} 

\begin{table*}
\begin{center}
  \caption{Initial compositions used in the stellar models.}.
\label{tab:initial}
\begin{tabular}{cccc} \hline \hline
Metallicity, $Z$   & Helium, $Y$  &  Hydrogen, $X$ & [Fe/H] \\ \hline
 0.04 &  0.330 & 0.630 & +0.46 \\
 0.05 &  0.354 & 0.596 & +0.55 \\
 0.06 &  0.375 & 0.565 & +0.63 \\
 0.07 &  0.396 & 0.534 & +0.70 \\
 0.08 &  0.420 & 0.500 & +0.66 \\
 0.09 &  0.438 & 0.472 & +0.81 \\
 0.10 &  0.459 & 0.441 & +0.85 \\ 
\hline \hline
\end{tabular}
\medskip\\
\end{center}
\end{table*}

We have calculated new stellar evolutionary sequences with initial masses between 1--8$\Msun$ with metallicities $Z=0.04$ to $Z=0.1$, with steps of $\Delta Z=0.01$.  These masses cover the full mass range of AGB behaviour, from the long-lived, low-mass AGB stars of 1$\Msun$ to models at or just beyond the CO-core limit ($\approx 8\Msun$). 

All evolutionary sequences are evolved from the zero age main sequence through to the end of the  thermally-pulsing AGB phase, or until convergence difficulties end
the calculations.  We do not evolve through to the post-AGB to the white dwarf phase except for a few cases \citep[for details of post-AGB evolution we refer to][]{miller16}.

Mass-loss at very high metallicities is highly uncertain.  We include mass-loss on the RGB for $M < 3\Msun$ and no RGB mass-loss for $M\ge 3\Msun$. The assumption of no mass-loss on the RGB
is based the short RGB lifetimes of intermediate-mass stars and typically lose very little mass ($\lesssim 0.05\Msun$) before reaching the AGB. For RGB mass-loss we use the Reimer's formula and adopt the free parameter $\eta = 0.477$ from \cite{mcdonald15}, noting this value was calibrated to globular clusters, systems more metal-poor than we consider here.  For the AGB phase we use the semi-empirical \citet{vw93} mass-loss rate, which was derived for AGB stars in the Milky Way and Magellanic Clouds. We come back to the discuss the effect of mass-loss in \S~\ref{sec:discuss}. 

We use the Mixing-length Theory (MLT) of convection and assume instantaneous mixing in all convective regions.  We set the mixing-length parameter $\alpha = 1.86$. The borders between radiative and convective regions in stellar interiors is notoriously difficult to model; here we use the search for a neutrally stable point from the formal Schwarzschild boundary \citep{lattanzio86,frost96}. This method has problems predicting carbon-stars of low enough initial mass to match observations and some overshoot beyond the formal border is needed \citep{kamath12,karakas10b,karakas16}. Based on predictions around solar metallicities \citep[e.g.,][]{cristallo09,karakas16}, the minimum mass for TDU at a given metallicity is likely to be $0.5\Msun$ lower than we predict here. In \S~\ref{sec:discuss} we experiment with simple overshoot in order to test this aspect, using the same method as \citet{karakas16}.

We use the C and N-rich low-temperature molecular opacity tables from \citet{marigo09}, noting that this is the first time that such opacity tables have been used for very metal-rich models of $Z=0.1$. Other input physics used in the calculations is exactly the same as in \citet{karakas14b}. 

\subsubsection{Initial composition}

We present the initial compositions of the Monash stellar models in Table~\ref{tab:initial}. Here we include the global metallicity, $Z$, the helium abundance, $Y$, the hydrogen abundance, $X$, where abundances are in mass fraction. We also show the approximate [Fe/H] ratio, where we assume a scaled-solar initial mixture. Here we assume the standard spectroscopic notation: [Fe/H] $= \log_{10}$(Fe/H)$_{\rm star} -\log_{10}$(Fe/H)$_{\odot}$, and abundances are by number.

For the initial compositions in Table~\ref{tab:initial}, we determine the initial helium abundance according to
\begin{equation}
Y  = Y_{\rm p} + \frac{D Y}{D Z} Z ,
\end{equation}
where $Y_{\rm p}$ is the primordial helium abundance, which we set to 0.2485 \citep{aver13} and $DY/DZ = 2.1$ \citep{casagrande11}. We provide a detailed discussion of these choices in \citet{karakas14b}.

The initial C, N and O abundances are scaled solar, where we adopt the proto-solar values from Table~5 of \citet{asplund09}. The initial C/O ratio = 0.55 (solar) in all models.  Observations of unevolved main sequence stars show that the
[C/Fe] and [O/Fe] ratios decrease to slightly sub-solar values of $-0.1$ and $-0.2$, respectively for [Fe/H] $\ge 0$, while the C/O ratio increases \citep{bensby06,nissen14,amarsi19}.  The study of \citet{nissen14} extends to metallicities of [Fe/H] $\approx +0.5$ and shows that the C/O ratio plateaus around $0.8$ for metal-rich stars. This trend is confirmed by Galactic chemical evolution models \citep{cescutti09,romano19,romano20,kobayashi11a,kobayashi20}. There is however large scatter in the data, with some metal-rich stars showing close to or around solar values for C/O.

%%%%%%%%%%%%%%%%%%%%%%%%%%%%%%%%%%%%%%%%%%%%%%%
\subsection{The MESA Stellar Models} \label{sec:mesa}
We compute a supplementary grid of stellar tracks with MESA version 11701 \citep{MESAIV} covering input metallicities of $Z=0.06$ through $Z=0.09$ at a resolution of $0.01$ and masses between 4 and 9 $M_{\odot}$ at a resolution of $1M_{\odot}$. We also compute a grid at solar composition ($Z=0.014, Y = 0.29$) for comparison.
The models adopt a uniform mixing length of $1.9H_{\rm p}$. A grid adopting a uniform, enhanced helium abundance of $Y=0.49$ is computed as well as a separate grid adopting the scaled helium abundances used in the Monash models (see Fig~\ref{fig:hrd-m6}). This latter grid is used to compare the codes on the HR diagram.
We use the \verb|photosphere_tables| option for atmospheric boundary conditions, the \citet{asplund09} abundances, the Reimers red giant mass loss scheme with the free parameter set to $\eta=0.477$, same as the Monash models. On the AGB we use the \citet{bloecker95} wind scheme with an efficiency parameter of $\eta = 0.05$. We invoke the \verb|co_burn_plus.net| nuclear reaction network, which includes the dominant burning chains required by the proton-proton chains, CNO cycle, helium burning, carbon/oxygen burning, and alpha capture up through silicon (\iso{28}Si). As the standard equation of state tables in MESA do not cover metallicities beyond $Z\approx0.092$, we cannot match the Monash calculations up to $Z=0.1$, but our grid is sufficient for detecting the lower carbon burning threshold.  
We compute two grids: one which adopts no convective overshoot (as in the Monash stellar models) and one in which a fixed convective overshoot of 0.014 pressure-scaled heights ($0.014 H_p$) is set uniformly for all burning regions. The models are terminated according to the condition \verb|stop_at_TP = .true.|, which halts the evolution once a thermal pulse is detected, or an age limit estimated from Table \ref{tab:modelsz04}--whichever occurs first.

\begin{figure}
\begin{center}
\includegraphics[width=0.95\columnwidth]{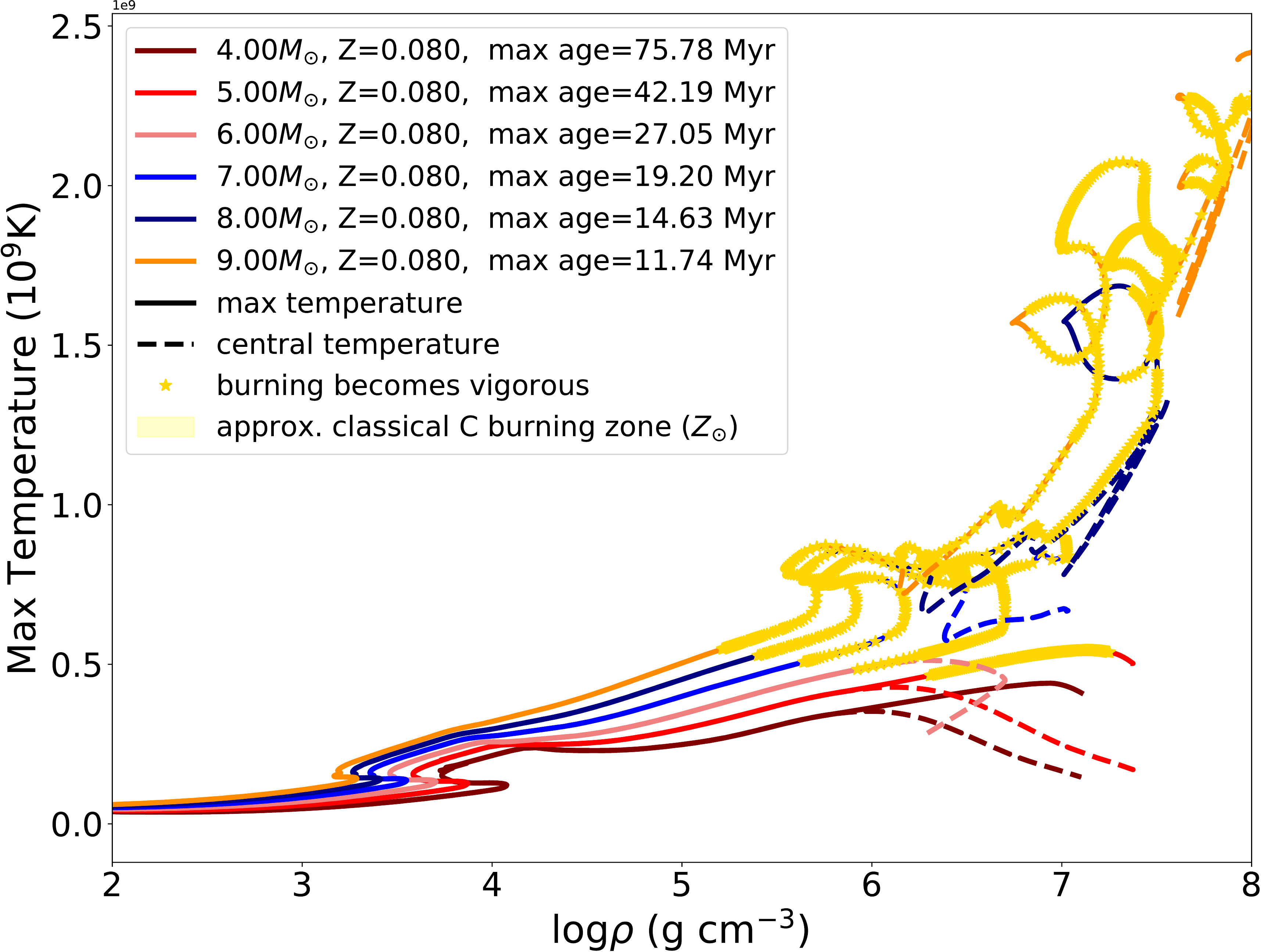}
 \caption{A $T_{\text{eff}}-\rho$ diagram is shown for a set of MESA models with varying mass and fixed composition $Z=0.08$ and $Y=0.49$. The rightward curling feature between $\log \rho =5.5$ to $\log \rho =7$ and $T_{\text{eff}}= 0.5 - 1.0 \times 10^{9}$K is characteristic of carbon burning. The yellow stars overlaid on the models indicate regions where burning is vigorous. The pale yellow rectangle highlights the approximate carbon burning regime for stars with a solar composition. Maximum ages for each model are indicated in the legend.}
\label{fig:MESA-T-rho}
\end{center}
\end{figure}

Detection of carbon ignition in the MESA models was performed by analysing Kippenhahn diagrams generated with a modified version of \verb|mkipp| \citep{marchantSW} and inspection of $T_\text{eff}-\rho$ diagrams, an example of which is shown in Figure \ref{fig:MESA-T-rho} for models of $Z=0.08$.

As in the calculations performed with the Monash code, we find that the carbon burning threshold drops in mass with increasing metallicity, with vigorous, sustained carbon burning appearing for masses as low as $5 M_{\odot}$ for the most metal-rich models ($Z=0.09$). 
Our findings for the grids with and without overshoot are summarised in Fig.~\ref{fig:cMESA}.

%%%%%%%%%%%%%%%%%%%%%%%%%%%%%%%%%%%%%%%%%%%%

\section{Results} \label{sec:results}

\begin{figure}
\begin{center}
\includegraphics[width=0.95\columnwidth]{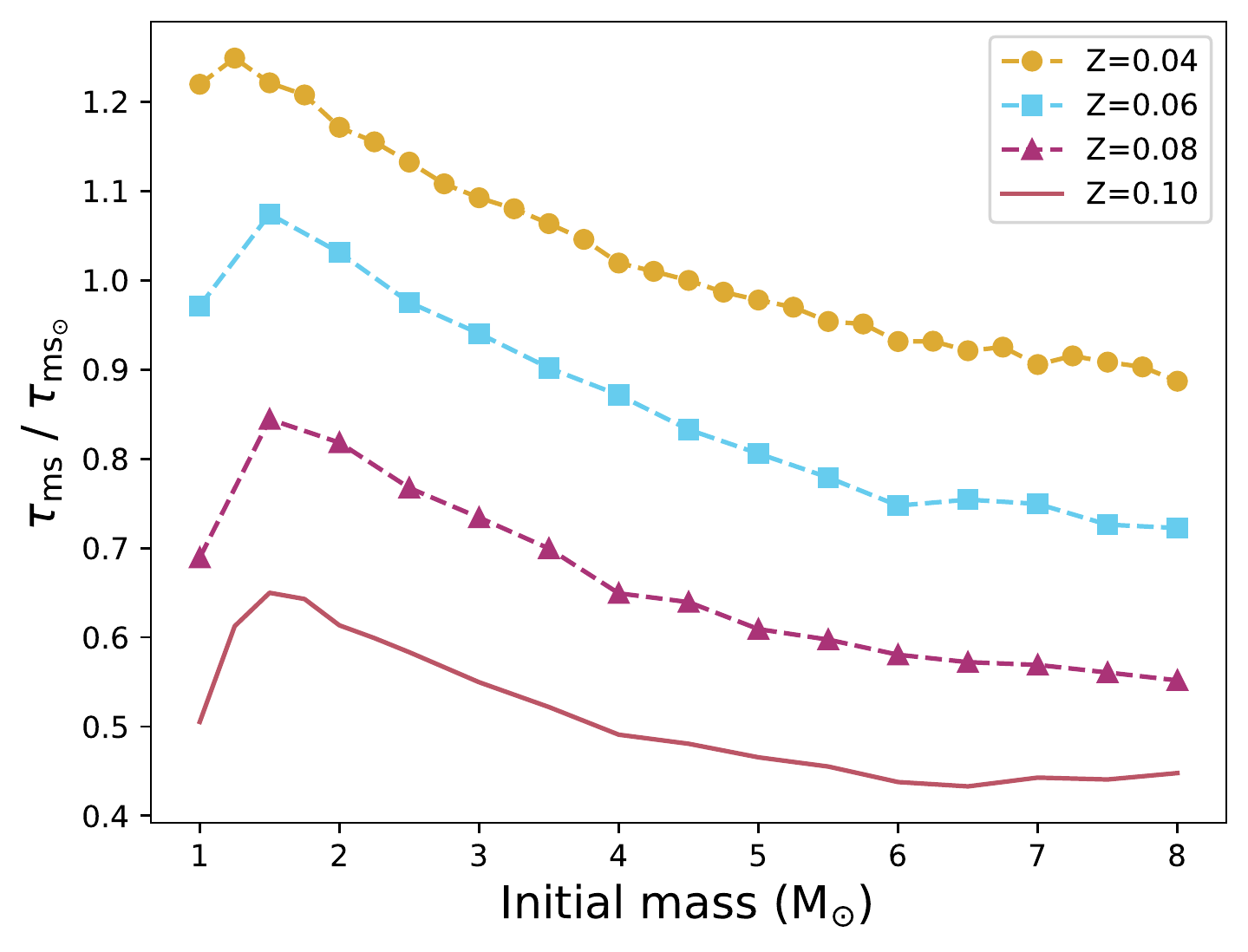}
 \caption{Ratio of main-sequence lifetime to the solar main-sequence lifetime for stellar models with $Z=0.04, 0.06,0.08$ and $Z=0.1$. 
 Solar-metallicity lifetimes are from \citet{karakas14b}.  \label{fig:tau-ms}}
\end{center}
\end{figure}

We start with stellar lifetimes in Fig.~\ref{fig:tau-ms}, where we show the ratio of the main-sequence lifetimes to the solar-metallicity main-sequence lifetime from the Monash models.
The main-sequence lifetimes of the $Z=0.04$ models are longer than the solar metallicity low-mass models $\lesssim 4\Msun$ and otherwise similar. Models of increasing metallicity show decreasing main-sequence lifetimes compared to the $Z=0.04$ models and by $Z=0.1$ the models show significantly shorter hydrogen-burning lifetimes. The $Z=0.1$ models show shorter lifetimes for two significant reasons: 1) the reduced hydrogen fuel available to them, and 2) most importantly, the increased luminosity on the main-sequence for each mass, as a consequence of a higher initial mean molecular weight, as discussed by \citet{meynet06}. To tease out the importance of increasing mean molecular weight we can examine the models of \citet{karakas14b}, which explored the effect of increasing the initial helium abundance in AGB models of the same metallicity. Changes in helium of $\Delta Y = 0.03$ cause the mean molecular weight to increase by 3\%, whereas the main-sequence lifetime changes by up to 20\%. However, increasing $\mu$ causes the central temperature to increase (from the perfect gas law) which consequently results in a higher rate of H-burning and a higher main-sequence luminosity.

The significant decrease in $\tau_{\rm ms}$ in the $Z=0.1$  model of $\approx 50$\% with respect to the solar metallicity model is consistent with results reported in \citet{mowlavi98b}.
The same relationship between increasing metallicity and decreasing main-sequence lifetime (and lifetime in general) is found in the MESA models. The maximum age of the MESA models--where ``max age'' is defined as the onset of thermal pulses, at which point the simulations are terminated--drops by at least two-thirds from solar composition ($Z=0.014$) to $Z=0.09$ across all masses considered. Lifetimes for the MESA models are shown in Figure \ref{fig:cMESA}.

\begin{figure}
\begin{center}
\includegraphics[width=0.95\columnwidth]{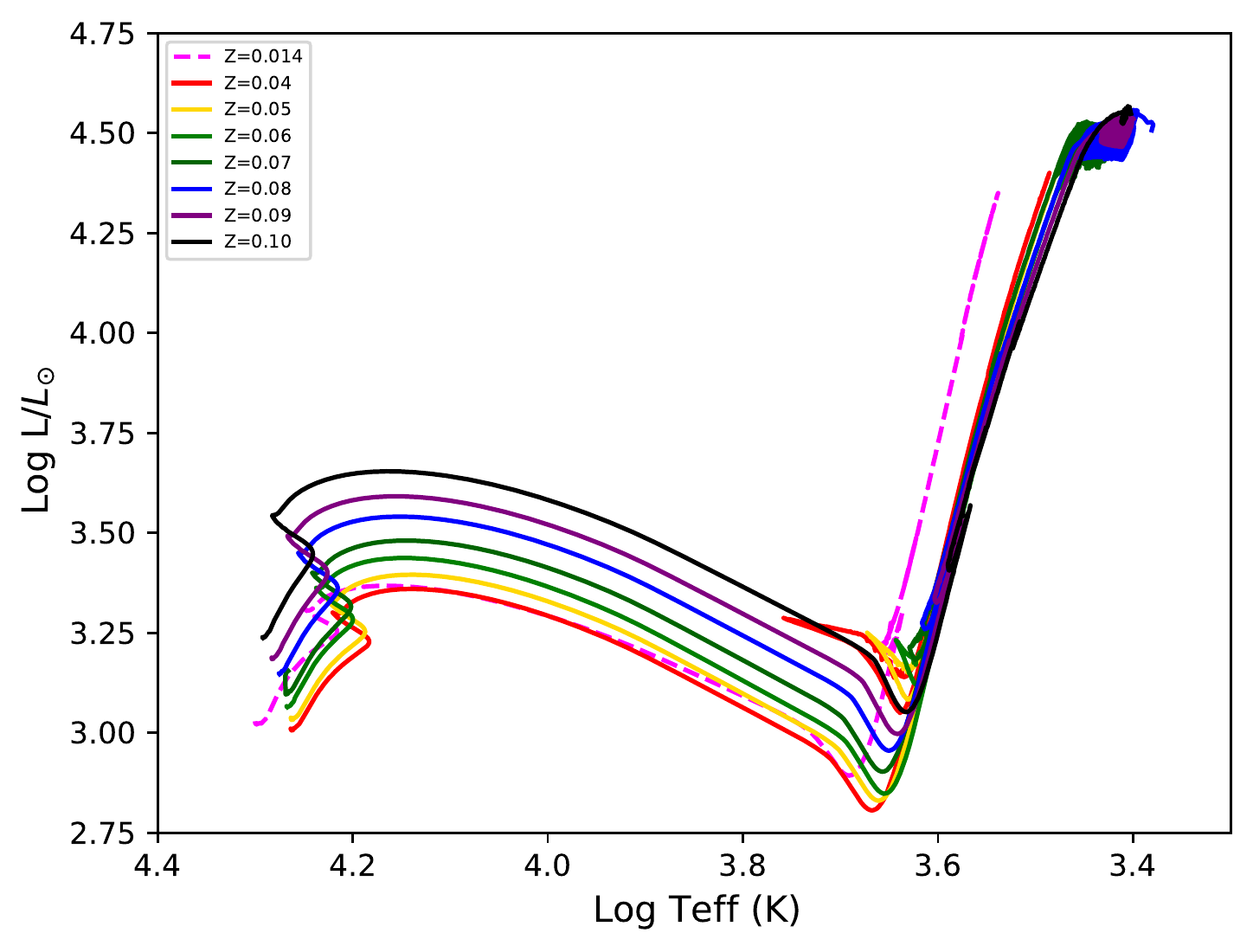}
\includegraphics[width=0.95\columnwidth]{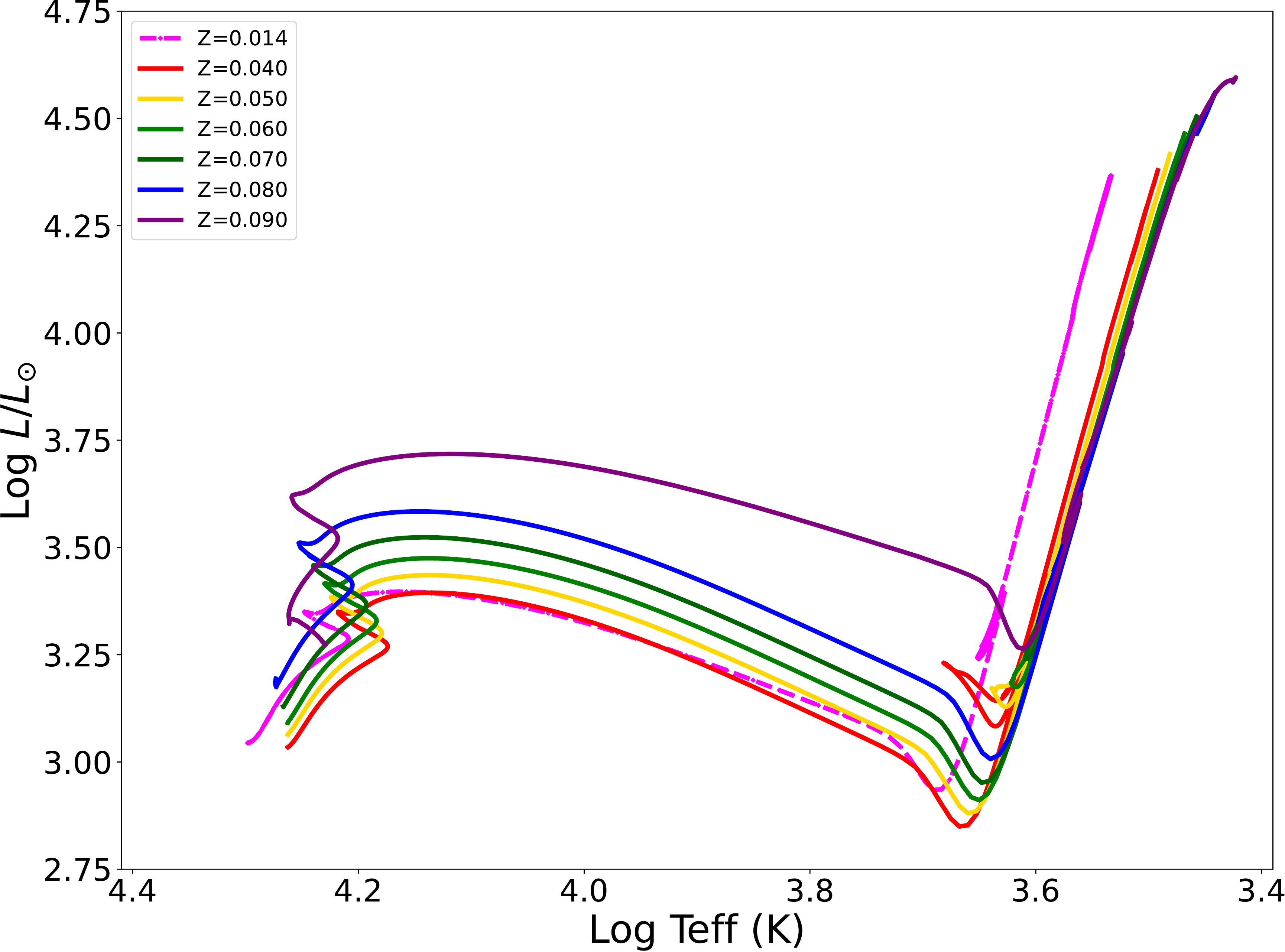}
 \caption{Evolutionary tracks for a selection of the 6$\Msun$ models from solar metallicity (dashed line, $Z=0.014$) to $Z=0.1$ from the Monash models (upper panel) and MESA models (lower panel, with maximum $Z=0.09$). \label{fig:hrd-m6}}
\end{center}
\end{figure}

In Fig.~\ref{fig:hrd-m6} we present stellar evolutionary tracks of models of 6$\Msun$ of various metallicity, evolved from the zero-age main sequence to the AGB phase, while the lower panel shows tracks from the MESA models. In the Appendix we include Hertzsprung-Russell diagrams (HRD) for models of $Z=0.04, 0.05, 0.06$, and $Z=0.1$ for all masses in Figs~\ref{fig:hrd-z04},~\ref{fig:hrd-z05},~\ref{fig:hrd-z06} and~\ref{fig:hrd-z10}. In Fig.~\ref{fig:hrd-m6} the thermally-pulsing AGB phase has been left out in most cases but can clearly be seen in the Monash evolutionary tracks of the 6$\Msun$, $Z=0.07$ and $Z=0.08$ models.
 
From Fig~\ref{fig:hrd-m6} and Fig.~\ref{fig:hrd-z10} (in the Appendix) we see some key features of models $Z \approx 0.1$ \citep[see discussions in][]{mowlavi98a,mowlavi98b,meynet06}: 1) the higher luminosities compared to their lower metallicity counterparts, as a result of higher central burning temperatures. This means that the 1$\Msun$, $Z=0.1$ model burns hydrogen predominantly via the CNO cycle and has a convective core on the main sequence, evidenced by the evolutionary track shown in Fig~\ref{fig:hrd-z10} \citep[e.g., as also found by][]{mowlavi98a,mowlavi98b}.
There is a transition metallicity for the development of a convective core in 1$\Msun$ models and that is $Z \ge 0.08$; models with metallicity below have radiative cores for most of the main sequence and have a similar structure to models of solar metallicity. An important consequence of the $Z=0.1$ models burning hotter and at higher luminosity on the main sequence is that they enter the post-main sequence phase with larger H-exhausted cores. 

Fig.~\ref{fig:hrd-m6} shows how increasing metallicity well above $Z = 2Z_{\odot}$ change the evolutionary behaviour of 6$\Msun$ theoretical models. The Monash models shown in the upper panel follow a fairly smooth transition from $Z=0.03$ \citep[using data from][]{karakas14b} to $Z=0.1$, with the solar metallicity track shown as a dashed-line. Models with $Z\le 0.05$ begin the main sequence cooler than the solar-metallicity model, whereas models with $Z>0.05$ are hotter and brigher than the solar metallicity model from the zero-age main sequence to the giant branch. By the time the stars reach the AGB the more metal-rich models are overall cooler, owing to a higher opacity, and also considerably brighter at the tip of the AGB, as a consequence of the increasing H-exhausted core mass.

The $6M_{\odot}$ MESA models, shown in the lower panel of Fig~\ref{fig:hrd-m6}, display similar behaviour with increasing metallicity when compared with the Monash models. The MESA models shown in Fig.~\ref{fig:hrd-m6} have been calculated with the same initial helium mass fraction as the Monash models, as shown in the figure legend, and as such show a similar transition to increasing luminosity with increasing metallicity. The one exception is the most metal-rich 6$\Msun$, $Z=0.09$ MESA model, which shows a much higher luminosity at the main-sequence turn-off and sub-giant branch, compared to the Monash models which show a smoother transition from $Z=0.08$ to $Z=0.1$. We speculate that the main reason for this abrupt increase in $L$ for the $Z=0.09$ MESA model may have to do with the opacity extrapolation to these highest metallicities.  The luminosities and temperatures of the MESA tracks at the tip of the AGB monotonically increase and decrease, respectively, as a function of metallicity. 

\subsection{First giant branch}

The ascent of the first giant branch (or RGB) is characterised by a deepening of the convective envelope and the first canonical mixing event, the first dredge-up (FDU). The second dredge-up (SDU) occurs later on, during the early AGB, after the model has exhausted its supply of helium in the core. The FDU and SDU mixing events are well studied in theoretical stellar evolutionary models, although most calculations only extend to about twice solar in terms of maximum metallicity
\citep[e.g.,][]{becker79, maeder89, eleid94, charbonnel96, forestini97, dominguez99,umeda99, siess06, doherty14a}.

Both FDU and SDU mixing events dredge-up the products of hydrogen burning from the deep interior to the stellar surface. The depth of mixing depends on the initial mass and metallicity, where FDU is deeper in lower mass stars of $\lesssim 4\Msun$ while SDU is deeper and reaches the products of complete H-burning in intermediate-mass stars $\gtrsim 4\Msun$. Here we comment on the depth of these mixing events, as a function of metallicity and mass for $Z\ge 0.04$.

\begin{figure}
\begin{center}
\includegraphics[width=0.9\columnwidth]{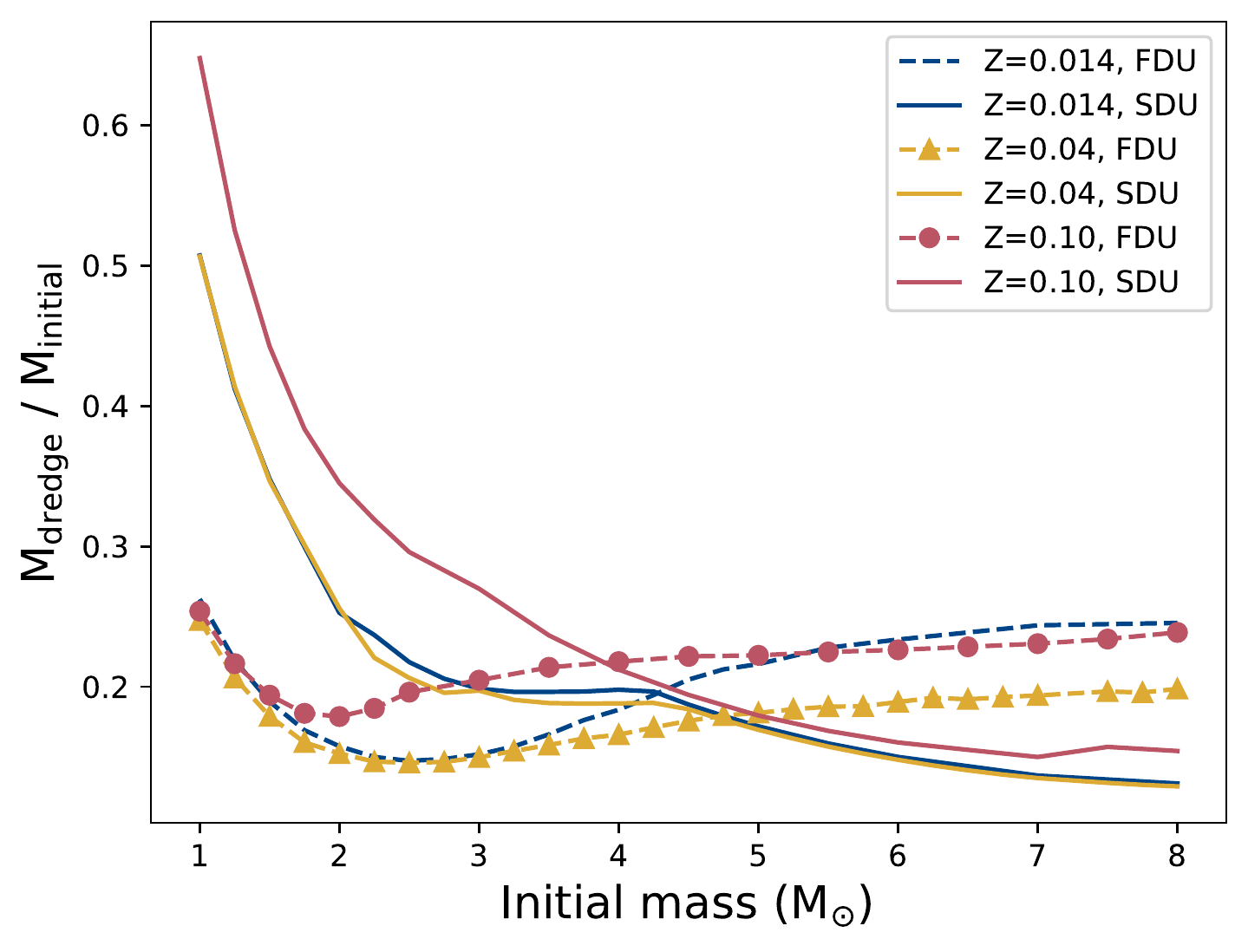}
 \caption{Depth of the first (dashed lines with symbols) and second dredge-up (solid lines) for models of solar metallicity ($Z=0.014$), $Z=0.04$ and $Z=0.1$. The depth is the ratio of the deepest inward extent in mass as a ratio of the initial stellar mass, as a function of initial mass.  \label{fig:depth}}
\end{center}
\end{figure}

In Figure~\ref{fig:depth} we show the depth of first and second dredge-up as a function of initial stellar mass for models of solar metallicity ($Z=0.014$), $Z=0.04$, and $Z=0.1$. We can compare this figure to the depth of FDU and SDU found in models of lower metallicities \citep[e.g.,][]{boothroyd99,karakas14dawes}. FDU in low-mass models of $Z=0.04$ are similar to their solar metallicity counterparts, while FDU is deeper in intermediate-mass $Z=0.04$ models. The increase in FDU depth (relative to the total stellar mass) with increasing metallicity for intermediate stellar masses is consistent with models by \citet{boothroyd99} when comparing results for $Z=0.004$ to solar.

Of interest are the $Z=0.1$ models, which have shallower FDU than solar models in the range between 2 to 5$\Msun$ but are otherwise similar.  The depth of second dredge-up seems relatively independent of metallicity, even for models of $Z=0.1$. Likewise the minimum mass for second dredge-up is only mildly metallicity dependent, hovering around $4\Msun$ for $Z=0.04-0.06$. The minimum mass is reduced in models of $Z=0.1$ to 3.5$\Msun$ as a consequence of a larger H-exhausted core mass post-main sequence. 

The radius along the first giant branch is an important quantity for binary star interactions and the survival rates of planet around evolved stars \citep{hurley00, hurley02,rasio96,villaver07}. This is because the interactions with a companion are more likely to occur when the most-massive star first expands to giant dimensions. We can examine how metallicity influences the radius along the first giant branch as a function of stellar mass by plotting the maximum radius at the tip of the RGB.

\begin{figure}
\begin{center}
\includegraphics[width=0.90\columnwidth]{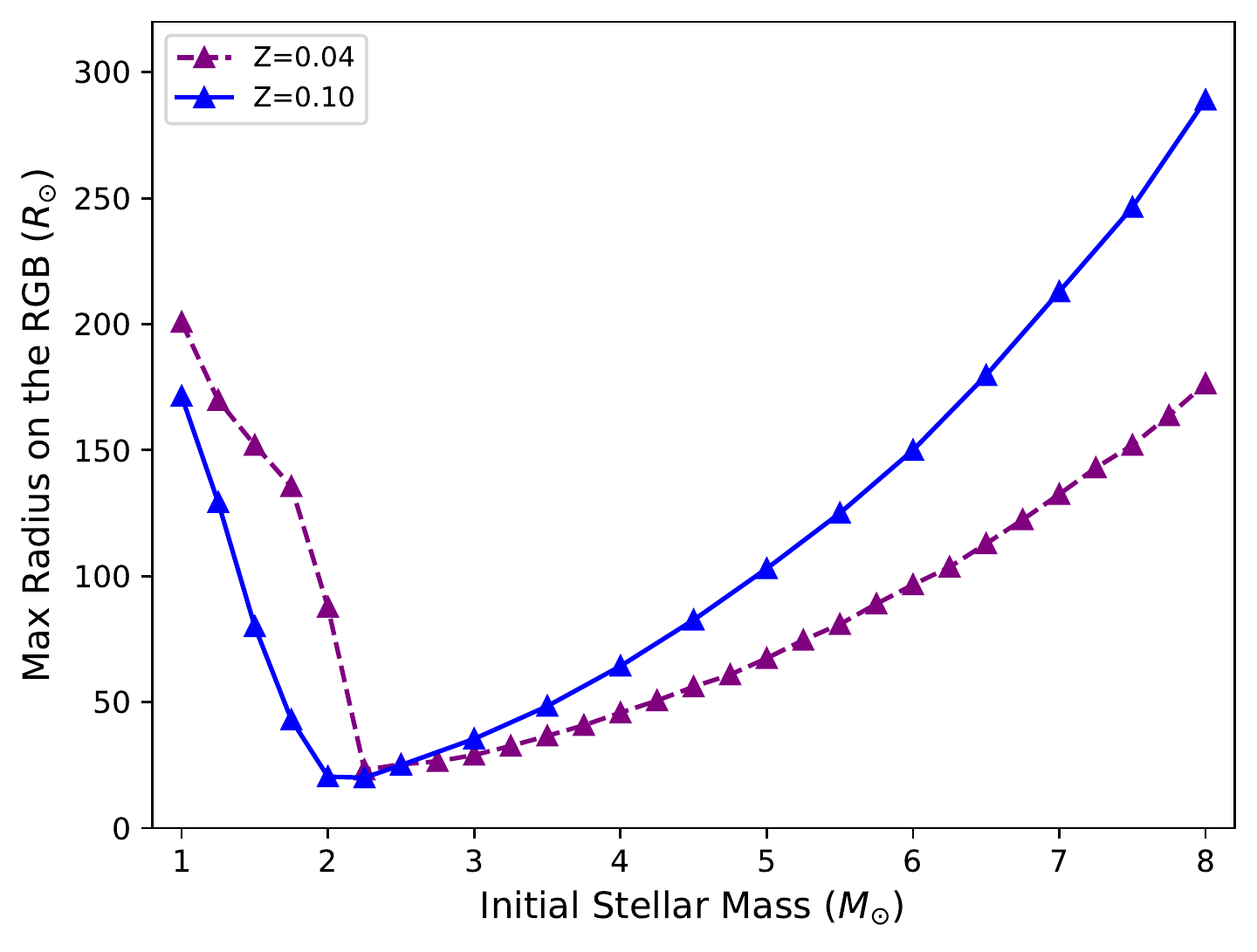}
\includegraphics[width=0.85\columnwidth]{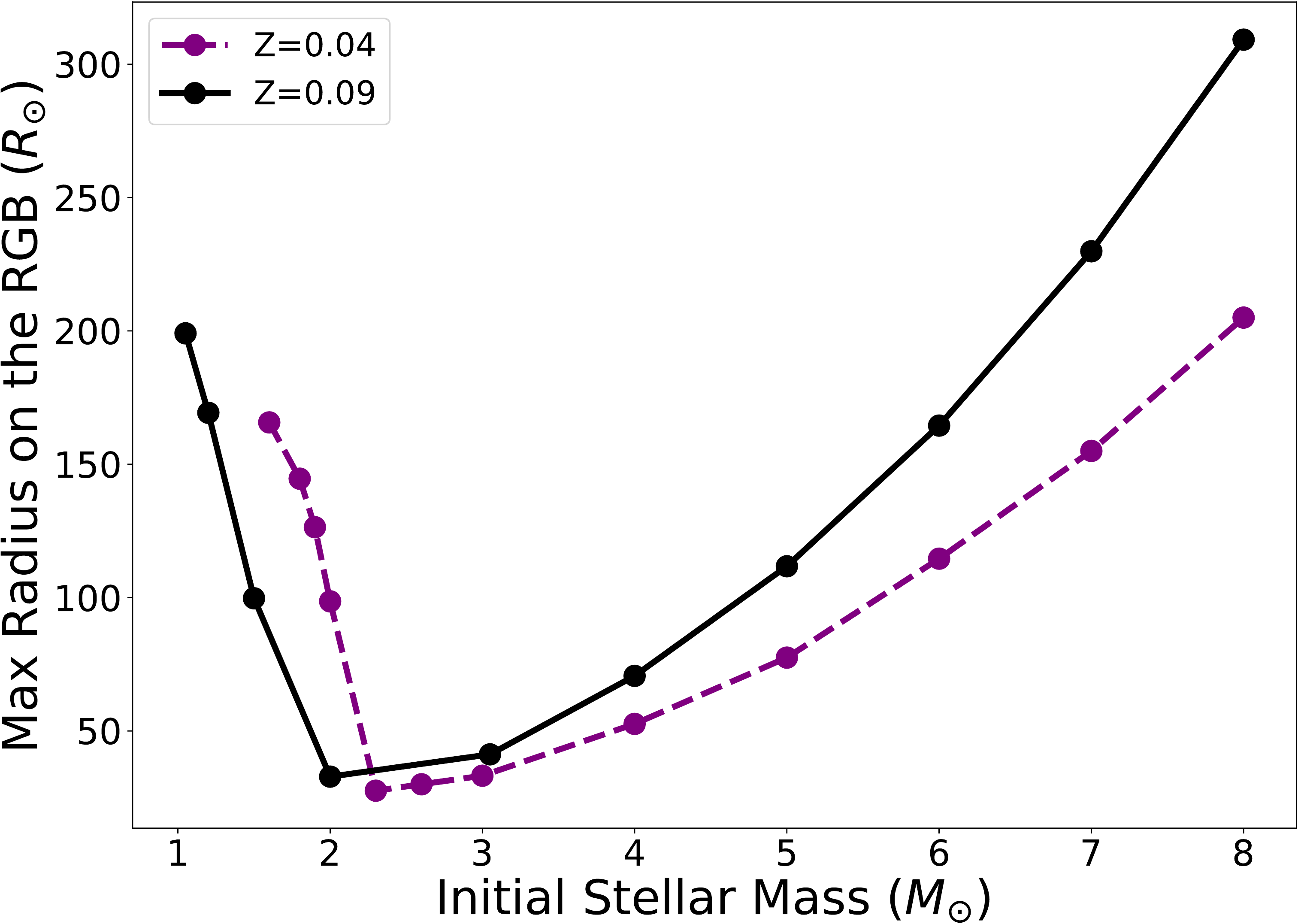}
 \caption{Top: The maximum radius (in $R_{\odot}$) on the red giant branch as a function of initial stellar mass, for models with $Z=0.1$ (black solid points and line) and $Z=0.04$ (blue open circles, dashed line). Bottom: The same for MESA models, with $Z=0.04$ in purple and $Z=0.09$ in black.
 \label{fig:maxr}}
\end{center}
\end{figure}

In the upper panel of Fig~\ref{fig:maxr} we show the maximum radius at the tip of the RGB for Monash models of $Z=0.04$ and $Z=0.1$. In the lower panel of Fig~\ref{fig:maxr}, we show the same for MESA models of $Z=0.04$ and $Z=0.09$.\footnote{Models with $Z=0.1$ lie outside of the standard opacity tables in MESA and hence produce convergence problems; we adopt $Z=0.09$ in lieu.} The radii at the tip of the RGB show similar trends with initial stellar mass, where the radii is larger in stars that experience the core He-flash, and shows a minimum at around 2$\Msun$, which is the transition from degenerate helium ignition to non-degenerate ignition. It can clearly be seen in Fig.~\ref{fig:maxr} that the minimum radii shifts to slightly higher mass when moving from models of $Z=0.04$ to the most metal-rich models shown ($Z=0.1$ for Monash; $Z=0.09$ for MESA). Across the two codes the radii are similar with differences of up to about 16\%, with the largest differences seen in models of around 8$\Msun$, and smaller differences of $\approx 10$\% seen in the lower mass stars, e.g., for models of 2$\Msun$, $Z=0.04$ the maximum radius is 88$\Rsun$ from Monash and 99$\Rsun$ from MESA.

Across codes, we find that the radius for low-mass stars that experience the core helium flash is directly related to their RGB lifetime; longer lifetimes result in higher tip RGB luminosities and radii. The main reason that the tip RGB radius correlates with stellar lifetime is related to the time taken for the cores of low-mass stars to reach the necessary temperatures for core He ignition. Lower mass stars take longer to reach temperatures of 100 million K and consequently their RGB lifetimes are longer; this equates to higher luminosities and cooler effective temperatures on the HRD and therefore larger radii. The core mass at helium ignition is roughly  $\approx 0.45\Msun$ for metallicities from solar to $Z=0.1$, although the core mass at the core flash does reduce with increasing metallicity, where e.g., at $Z=0.014$ the core mass is 0.46$\Msun$ compared to 0.42$\Msun$ for $Z=0.01$ (in models of 1$\Msun$).

The $Z=0.1$ and $Z=0.09$ models evolve off the main sequence with larger H-exhausted cores, so spend overall less time on the RGB (e.g., the 1.5$\Msun$, $Z=0.1$ model has a RGB lifetime that is 2.7 times shorter than the $Z=0.04$ model). Hence, the maximum radius at the tip of the RGB is larger in $Z=0.04$ models. In contrast, for intermediate-mass stars the maximum radius reached at the tip of the RGB depends primarily on initial composition, with the $Z=0.1$ and $Z=0.09$ models reaching larger radii owing to their high opacities as can be seen in Fig~\ref{fig:maxr}.

The maximum stellar mass for the core helium flash is dependent on mass, initial composition and the inclusion of convective overshoot on the main sequence \citep{bertelli86}. Here we include no overshoot in both the Monash and MESA models so the minimum mass for the core He-flash is higher compared to models with overshoot by $\approx 0.5\Msun$ \citep[e.g.,][]{pols98}. The maximum mass for core He-flash is about 2$\Msun$ for $Z=0.014$ \citep{karakas14b}, which rises to 2.5$\Msun$ by $Z=0.06$. For metallicities above this the trend is reversed and the minimum is around 1.75$\Msun$ for $Z=0.1$, similar to models of much lower metallicity around $Z\le 0.001$ \citep{fishlock14b}. \citet{mowlavi98b} find that the maximum mass for core He-flash is 1.5$\Msun$ for $Z=0.1$, lower than their models of solar metallicity. The reason for the reversal is clear: the most metal-rich models enter the RGB with larger core masses.

\subsection{AGB evolution} \label{sec:agb}

We now turn our attention to evolution during AGB. Following the exhaustion of helium in the core the convective envelope deepens once again, moving inwards and the star expands back to giant dimensions.
There are a number of reviews of AGB evolution, although none extend the discussion to the very metal-rich regime which we are studying here \citep[e.g.,][and references therein]{iben91,busso99,herwig05,karakas14dawes}.

\begin{figure}
\begin{center}
\includegraphics[width=0.95\columnwidth]{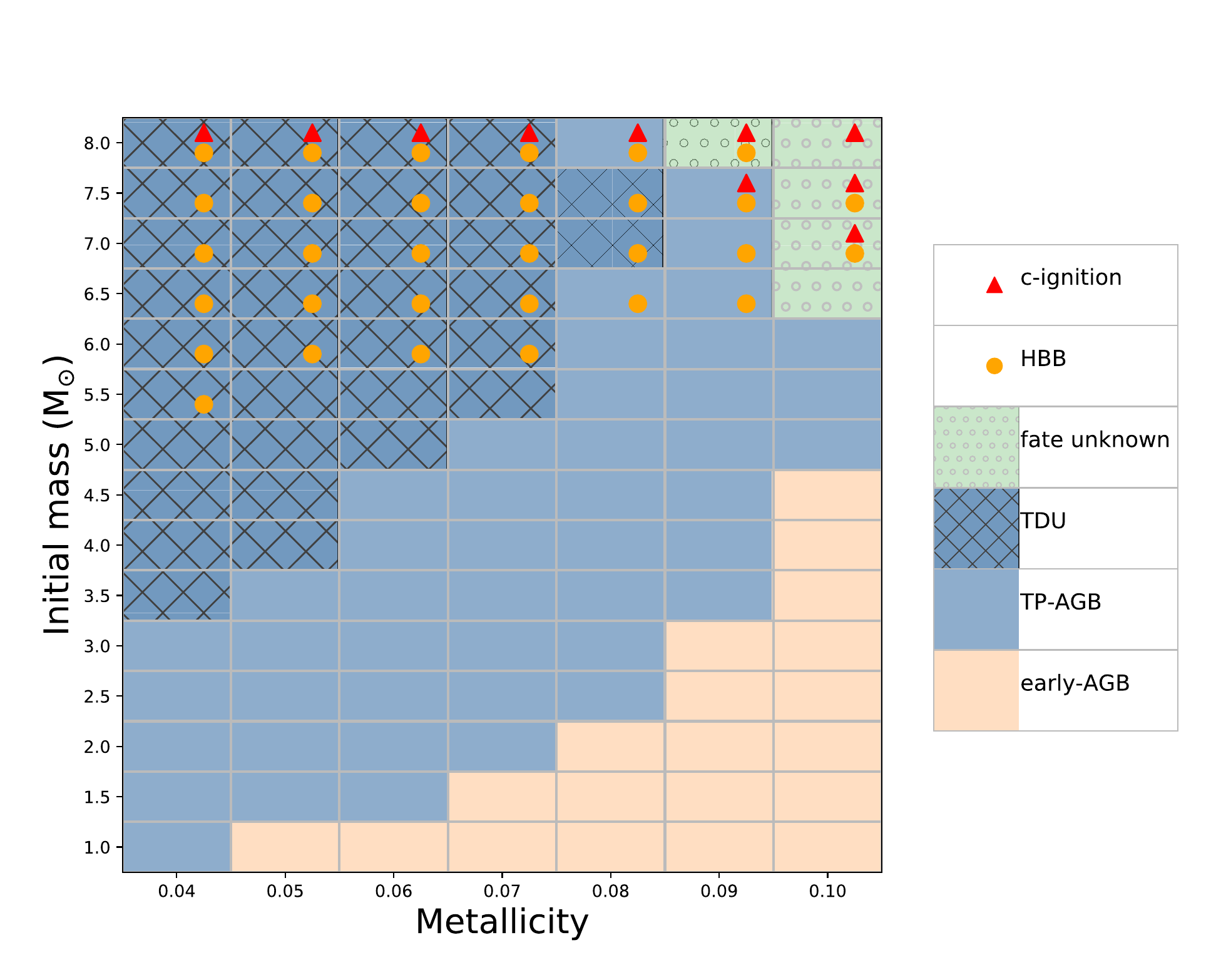}
 \caption{Pictorial view of the behaviour shown by the very metal-rich AGB models with mass-loss.
 Pale orange squares indicate models that exit the AGB phase as He stars before reaching the TP-AGB phase, pale blue are models that make it to the thermally-pulsing phase, and blue hashed boxes are models that also show TDU. Numerical difficulties ended the calculation of the models in green during SDU; these models would likely make it to the super-AGB phase. Models with HBB and carbon burning are also indicated. \label{fig:agb}}
\end{center}
\end{figure}

In Fig.~\ref{fig:agb} we illustrate the behaviour of the Monash AGB models as a function of mass and metallicity, where different coloured squares indicate if the model ends its life on the early-AGB, as a result of intense mass-loss, or makes it to thermally-pulsing AGB phase. We also indicate if the model experiences TDU, HBB and/or carbon burning. 

\subsubsection{Onset of carbon burning}

The maximum mass of a CO-core AGB star is defined by the onset of carbon burning. Stars between about 8 to 12$\Msun$ ignite carbon off centre before going through a brief phase of core carbon burning \citep{siess06,doherty10,jones13}. Stars at the lower end of this mass range (approximately 8-10$\Msun$) will enter the AGB with ONe cores and evolve through a super-AGB phase, before losing their envelopes through mass-loss to die as massive ONe white dwarfs. In contrast, C ignition moves progressively toward the centre in the upper mass range above about 10$\Msun$, and these stars may experience neon-burning flashes, before exploding as electron-capture supernovae \citep{jones14,doherty17}. The upper initial mass for a CO-core AGB star is about 8$\Msun$ for solar metallicity \citep{karakas14b,doherty14a}, although similar to the core He-flash the result depends on the inclusion of core overshoot during the main sequence and core He burning phases \citep{gilpons07}.

Another interesting feature of our very metal-rich models is that the onset of carbon burning moves to lower initial masses with increasing metallicity, as a consequence of earlier evolution. Among the Monash models, by $Z=0.1$ the minimum stellar mass to experience carbon ignition is the 7$\Msun$ model, which experiences off-centre carbon flashes, although carbon burning does not reach the centre. This is very similar to the behaviour experienced by the 8$\Msun$ model of solar metallicity \citep{karakas14b,doherty14a}. These models are not super-AGB stars although they are at the border, and would end their lives as hybrid C-O/O-Ne white dwarfs \citep{doherty17}. MESA models with $Z=0.06$ through $Z=0.08$  experience carbon ignition at $7 M_{\odot}$ as well, this drops to 6$\Msun$ for the case of $Z=0.09$. For MESA models that invoke a small amount of convective overshoot ($0.014 H_{p}$), this mass threshold drops even further to $6M_{\odot}$ at $Z=0.06 - 0.08$ and $5M_{\odot}$ at $Z=0.09$. The slight difference in carbon burning mass thresholds between the two codes is easily explained by differences in microphysics and convective prescriptions.

From Fig.~\ref{fig:agb} we see that all of the 8$\Msun$ Monash models with $Z$ $\ge 0.04$ experience carbon ignition. All of these models end their lives as hybrid white dwarfs, except the 8$\Msun$ models of $Z= 0.09$ and $Z=0.1$, which burn carbon through to the centre. Both of these are expected to become O-Ne white dwarfs, possibly after a super-AGB phase depending on the mass-loss rate. We did not follow the evolution of these models into the super-AGB phase, owing to numerical difficulties during carbon burning. We also did not follow the evolution of the 7$\Msun$, $Z=0.1$ model beyond SDU. The model had already lost just over 1$\Msun$ by the time the calculation ended, and would likely enter the thermally-pulsing phase, albeit at very high luminosity ($\log L/\Lsun \approx 4.68$). 

The analogue of Fig~\ref{fig:agb} for the MESA models is given in Fig~\ref{fig:cMESA}. This table provides the detection of carbon burning and other qualitative features of the stellar interior (e.g. location of the carbon burning region; whether or not there is a deeper convective pocket of burning present) as a function of mass and metallicity for MESA models with no convective overshoot. The criterion ``burning is vigorous'' refers to the detection of the 
\iso{12}C $+$ \iso{12}C burning chain coincident with neutrino losses, as in \citet{FarmerFieldsTimmes2015}. MESA models are uniformly halted at the beginning of the TP-AGB, so ``maximum age'' refers to the lifetime of the star from the ZAMS (plus an additional $\sim10^{-5}$ Myr, accounting for pre-main sequence convergence) to the onset of thermal pulses.

\citet{mowlavi98b} evolved models of 7$\Msun$ and 9$\Msun$ for $Z=0.1$ and note that the 9$\Msun$ experiences carbon burning. Inspection of the $\log \rho_{\rm c}-\log T_{\rm c}$ plot from that paper (their Fig.~3) shows that the 7$\Msun$ evolution may have ended during C ignition, with an upturn in temperature evident in the figure. The final position of the 7$\Msun$ is in a region that looks occupied by other models that experience C burning, consistent with our results.

\subsubsection{Thermally-pulsing AGB phase}

In this section we present results from the TP-AGB phase of evolution and concentrate our discussion from here on results from the Monash stellar models.   

In Table~\ref{tab:modelsz04} we present data for the AGB models of $Z=0.04$, with tables for the other metallicities included in the Appendix in Tables~\ref{tab:modelsz06-z07} and~\ref{tab:modelsz08-z10}. Each table contains the initial mass, if SDU occurs, the number of thermal pulses (TPs), the maximum TDU efficiency parameter, $\lambda_{\rm max}$, the minimum mass for the occurrence of thermal pulses (in $\Msun$), the maximum temperature at the base of the convective envelope (in $10^{6}$ K, MK), the maximum radiated luminosity (in the format $\log L_{\rm agb}^{\rm max}$, where $L$ is in $\Lsun$), the maximum luminosity from the He-shell (in the format $\log L_{\rm agb}^{\rm max}$, where $L$ is in $\Lsun$), the total stellar lifetime ($\tau_{\rm stellar}$ in Myr), and the RGB and AGB lifetimes, respectively ($\tau_{\rm rgb}$ and $\tau_{\rm agb}$, in Myr). The AGB lifetime includes time spent on the early AGB, before the first thermal pulse, along with the thermally-pulsing AGB lifetime. Table~\ref{tab:modelsz04} for the $Z=0.04$ models also includes the TP-AGB lifetime ($\tau_{\rm tp-agb}$, in Myr), which is the time from the first thermal pulse to the end of the AGB phase.
We define He-shell flashes (thermal pulses) to occur if the He-shell luminosity exceeds $\log L/\Lsun = 4$ and for the He-shell luminosity to exceed the radiated luminosity.

\begin{table*}
\begin{center}
\caption{Stellar models calculated with $Z=0.04$.}
\label{tab:modelsz04}
\begin{tabular}{lccccccccccc} \hline \hline
Mass & SDU & \#TP & $\lambda_{\rm max}$ & $M_{\rm c}(1)$ & 
$T_{\rm bce}^{\rm max}$ & $\log L_{\rm agb}^{\rm max}$ & 
$\log L_{\rm He-shell}^{\rm max}$ & $\tau_{\rm stellar}$ & 
$\tau_{\rm rgb}$ & $\tau_{\rm agb}$ & $\tau_{\rm tp-agb}$ \\
($\Msun$) &  &  &  & ($\Msun$) & (MK) & ($\Lsun$) & 
($\Lsun$) & (Myr) & (Myr) & (Myr) & (Myr) \\ \hline
\multicolumn{11}{c}{$Z=0.04$, $Y=0.330$ models.} \\ \hline
1.00 & No  &  6 & 0.00 & 0.573 & 1.13 & 3.67 & 5.12 & 15200 & 3347 & 23.6 & 0.26 \\
1.25 & No  &  7 & 0.00 & 0.580 & 2.13 & 3.73 & 5.24 & 6620 & 1811 & 30.0 & 0.28 \\
1.50 & No  &  8 & 0.00 & 0.582 & 2.62 & 3.75 & 5.27 & 3452 & 843 & 20.8 & 0.31 \\
1.75 & No  & 12 & 0.00 & 0.581 & 2.93 & 3.82 & 5.56 & 2076 & 376 & 19.0 & 0.49 \\
2.00 & No  & 17 & 0.00 & 0.577 & 3.26 & 3.90 & 5.89 & 1394 & 169 & 22.7 & 0.75 \\
2.25 & No  & 23 & 0.00 & 0.571 & 3.64 & 3.96 & 6.21 & 1132 & 68.0 & 25.3 & 1.1 \\
2.50 & No  & 26 & 0.00 & 0.579 & 4.05 & 4.02 & 6.37 & 851.2 & 38.8 & 21.7 & 1.1 \\
2.75 & No  & 28 & 0.00 & 0.584 & 4.58 & 4.07 & 6.51 & 645.0 & 25.8 & 17.9 & 1.0 \\
3.00 & No  & 28 & 0.00 & 0.610 & 5.28 & 4.11 & 6.65 & 493.1 & 18.2 & 13.1 & 0.90 \\
3.25 & No  & 27 & 0.35 & 0.634 & 6.20 & 4.15 & 6.94 & 443.5 & 13.2 & 11.1 & 0.74 \\
3.50 & No  & 22 & 0.58 & 0.672 & 7.50 & 4.19 & 7.28 & 302.0 & 10.2 & 8.29 & 0.49 \\
3.75 & No  & 18 & 0.67 & 0.715 & 10.0 & 4.22 & 7.39 & 244.2 & 8.00 & 6.25 & 0.31 \\
4.00 & No  & 16 & 0.71 & 0.761 & 14.0 & 4.27 & 7.36 & 203.1 & 6.37 & 4.61 & 0.19 \\
4.25 & Yes & 14 & 0.74 & 0.809 & 17.1 & 4.32 & 7.28 & 167.4 & 5.23 & 3.86 & 0.12 \\
4.50 & Yes & 15 & 0.77 & 0.835 & 21.5 & 4.35 & 7.32 & 141.6 & 4.29 & 3.09 & 0.10 \\
4.75 & Yes & 18 & 0.80 & 0.844 & 27.6 & 4.38 & 7.47 & 121.1 & 3.61 & 2.61 & 0.11 \\
5.00 & Yes & 21 & 0.85 & 0.854 & 35.8 & 4.40 & 7.64 & 106.1 & 3.04 & 2.18 & 0.12 \\
5.25 & Yes & 24 & 0.86 & 0.863 & 46.6 & 4.42 & 7.70 & 92.5 & 2.51 & 1.88 & 0.12 \\
5.50 & Yes & 28 & 0.87 & 0.872 & 56.3 & 4.45 & 7.77 & 82.0 & 2.22 & 1.60 & 0.13 \\
5.75 & Yes & 30 & 0.87 & 0.883 & 61.8 & 4.48 & 7.79 & 73.4 & 1.82 & 1.38 & 0.12 \\
6.00 & Yes & 32 & 0.87 & 0.895 & 65.0 & 4.51 & 7.74 & 65.3 & 1.57 & 1.21 & 0.12 \\
6.25 & Yes & 35 & 0.88 & 0.906 & 67.8 & 4.54 & 7.80 & 59.1 & 1.31 & 1.08 & 0.12 \\
6.50 & Yes & 37 & 0.87 & 0.920 & 70.6 & 4.56 & 7.75 & 54.3 & 1.15 & 0.90 & 0.11 \\
6.75 & Yes & 39 & 0.86 & 0.934 & 73.2 & 4.59 & 7.70 & 49.1 & 0.95 & 0.82 & 0.10 \\
7.00 & Yes & 43 & 0.85 & 0.951 & 76.2 & 4.62 & 7.68 & 44.7 & 0.82 & 0.69 & 0.091 \\
7.25 & Yes & 45 & 0.85 & 0.970 & 79.2 & 4.65 & 7.66 & 41.6 & 0.72 & 0.62 & 0.082 \\
7.50 & Yes & 50 & 0.85 & 0.993 & 82.8 & 4.68 & 7.56 & 38.7 & 0.63 & 0.52 & 0.074 \\
7.75 & Yes & 56 & 0.85 & 1.014 & 86.2 & 4.71 & 7.55 & 36.1 & 0.55 & 0.47 & 0.070 \\
8.00$^{\rm a}$ & Yes & 64 & 0.84 & 1.039 & 89.6 & 4.76 & 7.69 & 32.7 & 0.53 & 0.45 & 0.064 \\
\hline \hline
\end{tabular}
\medskip\\
\noindent (a) Model ignited carbon on early AGB, with the core temperature nearing 900~MK.
\end{center}
\end{table*}

From Tables~\ref{tab:modelsz04},~\ref{tab:modelsz06-z07} and~\ref{tab:modelsz08-z10} we notice two things: 1) the peak flash intensity as measured by the maximum He-shell luminosity is reduced in models of increasing metallicity, for a given mass; and 2) the number of thermal pulses strongly decreases with increasing metallicity. The decreasing strength of thermal pulses with metallicity has important implications on the efficiency of TDU in these highest metallicity AGB stars, which is discussed in Section~\ref{sec:tdu}. 
The reduction in the number of thermal pulses is mostly as a result of increasing mass-loss, driven by increasing radius and luminosity on the AGB with increasing $Z$. This has implications for the mass-loss rate, especially on the early part of the AGB before thermal pulses begin. We use the AGB mass-loss rate from \citet{vw93}, where the mass-loss rate on the early part of the AGB is determined by the 
radial pulsation period, which depends on radius and mass according to
\begin{equation}
    \log P/{\rm days} = -2.07 + 1.94 R/\Rsun - 0.9 M/\Msun.
\end{equation}
However, once the period exceeds 500 days a luminosity driven superwind begins, where the mass-loss rate is proportional to $L$. 
Owing to their larger radii, very metal-rich models enter the AGB with larger pulsation periods and higher mass-loss rates than models with smaller $Z$. For example, the 2$\Msun$, $Z=0.04$ model has a period of $\approx 120$~days at the start of the AGB, much shorter than the $\approx 300$~days of the $Z=0.06$ model of the same mass. The main consequence is that the envelope mass is stripped more quickly in models with increasing $Z$, which reduces the number of thermal pulses, or, in some instances, stops them completely.

Given the above criteria we set for the occurrence of thermal pulses (e.g., the He-shell luminosity must exceed $\log L/\Lsun \ge 4$ and the He-shell luminosity must exceed the radiated luminosity) most of the $Z=0.1$ models do not experience a thermally-pulsing AGB phase. This is because of the intense mass-loss during the early AGB. Some of the $Z=0.1$ models that do not experience thermal pulses show very small He-shell luminosity oscillations (e.g., the 4$\Msun$, 4.5$\Msun$ and 6.5$\Msun$, $Z= 0.1$ models). In these cases the He-shell luminosity always remains below the radiated luminosity and true He-shell flashes do not develop. The $Z=0.1$ models that do develop thermal pulses experience much weaker He-shell flashes than than their lower metallicity counterparts, with $\log L(\Lsun) < 5$ for all models. A different choice of AGB mass-loss rate that results in weaker mass-loss during the early AGB may lead to these stars experiencing stronger thermal pulses.

Other models to skip the thermally-pulsing phase are shown in Fig.~\ref{fig:agb} and include all 1$\Msun$ models with $Z\ge 0.05$. These models end their lives as stripped He-stars and evolve toward the blue on the HR diagram at lower luminosities than is typical for post-AGB stars ($\approx 10^{4}\Lsun)$ but higher than the post-RGB stars found in the Magellanic Clouds \citep[e.g.,][]{kamath15}. For example, the 1$\Msun$, $Z = 0.06$ exits the AGB at around $\log L/\Lsun \approx 3.5$. The luminosity slowly drops to $\log L/\Lsun \approx 3$ as the the model moves toward hotter effective temperatures of 100,000~K, at which point the model evolves toward the white dwarf cooling track. 

Binary evolution can produce stripped He stars, where mass transfer removes the H-rich envelope. This behaviour is not expected from single stellar evolution as a consequence of mass-loss but very metal-rich models experience higher mass-loss rates owing to their high opacity envelopes. \citet{mowlavi98a} found their massive models $\gtrsim 60\Msun$ become helium stars as a consequence of mass-loss, similar to the behaviour we find for most of the $Z=0.1$ models. The main uncertainty around these predictions is the rate of mass-loss in evolved stars, which is unknown for these very high metallicities.
We explore the uncertainty around mass-loss on metal-rich AGB models further in Section~\ref{sec:discuss}.

\subsubsection{Third dredge-up} \label{sec:tdu}

The efficiency of TDU is known to strongly depend on stellar metallicity for a given mass \citep[e.g.,][]{boothroyd88c,karakas02}. The effiency of TDU can be quantified by the parameter $\lambda$, which is defined as follows:
\begin{equation}
    \lambda = \frac{\Delta M_{\rm dredge}}{\Delta M_{\rm h}},
\end{equation}
where $\Delta M_{\rm dredge}$ is the amount of material dredged into the envelope and $\Delta M_{\rm h}$ is the amount by which the H-exhausted core mass grew during the preceding interpulse period. The parameter $\lambda$ in theoretical calculations also varies with initial mass for a given $Z$, and time on the TP-AGB phase. For models that experience TDU, $\lambda$ increases from zero to reach an asymptotic value after some number of thermal pulses \citep{karakas02}. The strength of thermal pulses is directly connected to the efficiency of TDU, hence lower metallicity models, with stronger He-shell flashes, have deeper TDU \citep{marigo99}, while metal-rich models consequently should experience shallow TDU owing to weaker thermal pulses. In \citet{karakas14b} we examined how $\lambda$ and the minimum core mass for TDU varied in models of $Z=0.014$ to $Z=0.03$, noting how the metal-rich models experience shallow TDU at a given mass, or none at all. 

The minimum initial mass for TDU in solar metallicity AGB models is around 2$\Msun$ without any convective boundary mixing, which increases to 2.5$\Msun$ by $Z=0.03$ \citep{karakas14b,karakas16}. Increasing the metallicity still higher, we find that the minimum mass is 3.25$\Msun$ for $Z=0.04$, 4$\Msun$ for $Z=0.05$,  5$\Msun$ for $Z=0.06$, 5.5$\Msun$ for $Z=0.07$ and 7$\Msun$ for $Z=0.08$. These boundaries are illustrated in Fig~\ref{fig:agb}, which shows the parameter space of the models to experience TDU events. In summary, models below 3$\Msun$ do not experience TDU for $Z\ge 0.04$ and above $Z=0.08$, TDU disappears completely, even for the most massive AGB models. The disappearance of TDU in the most metal-rich models is likely caused by the weak thermal pulses experienced by these models.

\begin{figure}
\begin{center}
\includegraphics[width=0.9\columnwidth]{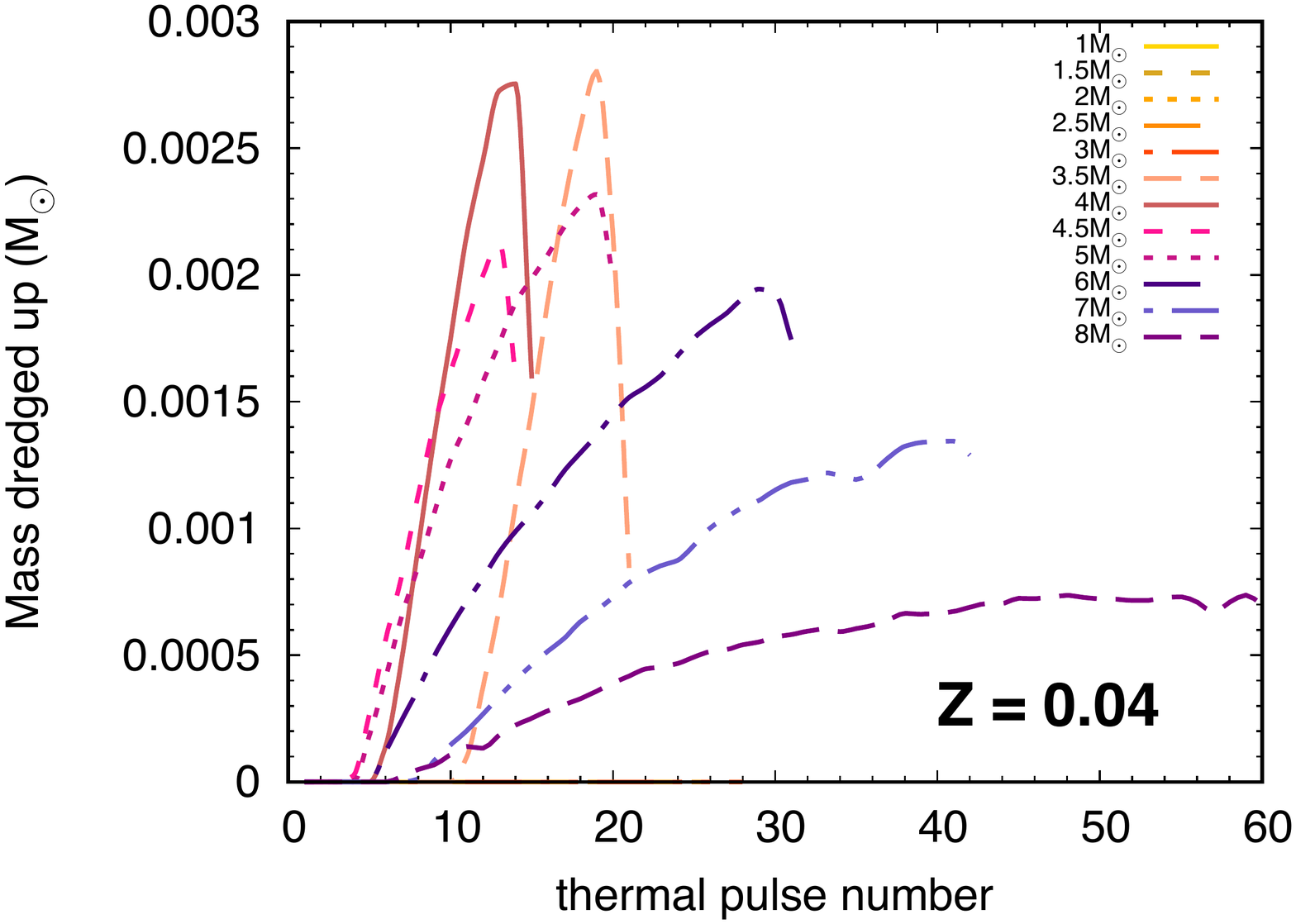}
\includegraphics[width=0.9\columnwidth]{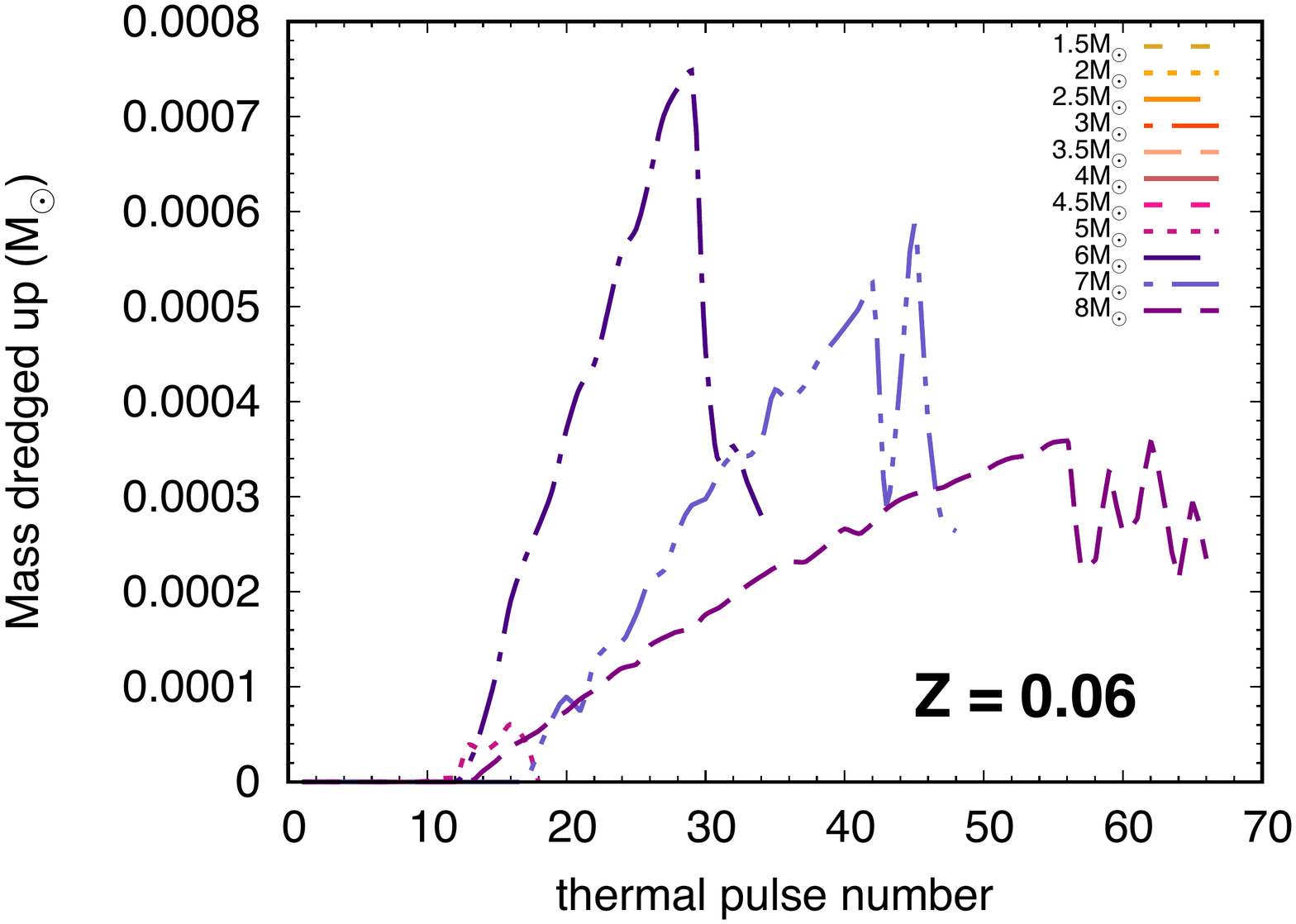}
 \caption{Mass dredged into the convective envelope as a function of thermal pulse number, for AGB models of $Z=0.04$ (upper panel) and $Z=0.06$ (lower panel).  Models not visible have no TDU.} \label{fig:mdredge}
\end{center}
\end{figure}

We present the maximum TDU efficiency parameter for all of the stellar models in Tables~\ref{tab:modelsz04},~\ref{tab:modelsz06-z07} and~\ref{tab:modelsz08-z10}. In Fig.~\ref{fig:mdredge} we show the mass dredged into the convective envelope by TDU as a function of thermal pulse number, for models of $Z=0.04$ and $Z=0.06$ from 1$\Msun$ to 8$\Msun$. Models not visible in Fig.~\ref{fig:mdredge} have no TDU. This plot is instructive because it shows two things:  that only intermediate-mass models over 5$\Msun$, $Z=0.06$ experience TDU and that the amount of mass dredged into the envelope at a given initial mass is also lower, by almost a factor of 3. We can quantify this more directly by summing the mass dredged into the envelope over the AGB for models of 5$\Msun$ with $Z=0.04$ and $Z=0.06$. The $Z=0.04$ model dredges up 0.024$\Msun$ while the $Z=0.06$ dredges up in total only $2.3\times 10^{-4}\Msun$, or about 100 times less. The main consequence will be for the stellar yields of metal-rich AGB stars, which we explore in a companion paper (Cinquegrana \& Karakas 2021, MNRAS, submitted).

In \citet{karakas14b} we compared initial mass ranges for carbon stars, noting that around solar metallicity the range extended from about 2$\Msun$ to 4.5$\Msun$, whereas by $Z=0.03$ the mass range for C-star production had shrunk to 3-4$\Msun$. The formation of a C-rich atmosphere has important consequences for mass-loss and dust production, as has been outlined in many papers \citep{nanni13,dellagli17,dellagli19,hoefner18}. None of our models become carbon rich on the AGB, where the maximum C/O $\approx 0.5$ is found in models around 3.5-4$\Msun$, $Z= 0.04$, which is an increase from the first dredge-up value of C/O $\approx 0.35$ (c.f. the initial C/O = 0.55 in all models). At higher metallicities still, the minimum mass for TDU is pushed to higher masses which means two things. One, the larger envelope means that any He-shell material is diluted over a larger mass, and second, the mass range that experiences TDU is almost the same that experiences HBB (e.g., see Fig.~\ref{fig:agb}), which keeps the C/O ratio low owing to CNO cycling. For metallicities above $Z=0.04$, the C/O ratio does not exceed 0.4. 

The numerical treatment of convective borders is a major uncertainty in theoretical models of AGB stars. We use the algorithm described by \citet{lattanzio86} to search for a neutrally stable point for the border between convective and radiative zones. This scheme has been shown to increase the amount of TDU compared to models that define the border according to the formal Schwarzschild criteria \citep{frost96,mowlavi99a}. Even with this scheme, Monash models have trouble matching observations of low-mass C-rich stars \citep[e.g.,][]{karakas10b,kamath12}, which has necessitated the need for some form of mixing beyond the formal border, at least for the lowest mass stars to experience TDU. We explore these issues are explored further in Section~\ref{sec:discuss}.

\subsubsection{Hot bottom burning}

Stars over about 4$\Msun$ experience HBB on the AGB, which occurs when the base of the envelope becomes hot enough for proton-capture nucleosynthesis. From a structural point of view, the main change is an increase in the radiated luminosity, which causes deviations from the classical core-mass luminosity relationship \citep[e.g.,][]{bloecker91,lattanzio92,boothroyd92}. The base of the envelope is less dense than the H-shell which means that higher minimum temperatures are required for CNO cycling than found typically in the cores of intermediate-mass stars on the main sequence (e.g., $T \gtrsim 20 \times 10^{6}$~K). We define models to have HBB if the temperature at the base of the envelope exceeds $T_{\rm bce} \ge 50 \times 10^{6}$~K.

\begin{figure}
\begin{center}
\includegraphics[width=0.9\columnwidth]{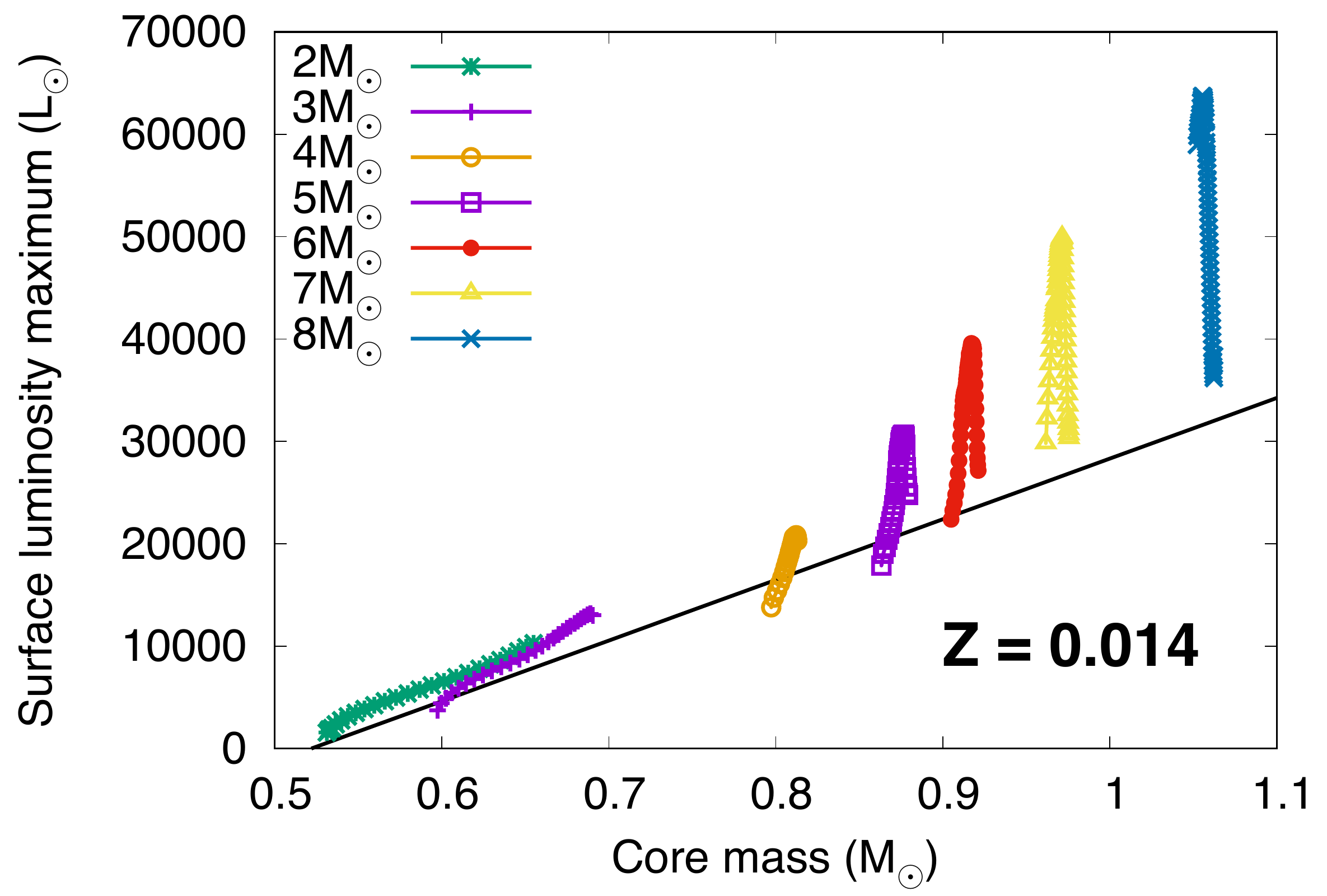}
\includegraphics[width=0.9\columnwidth]{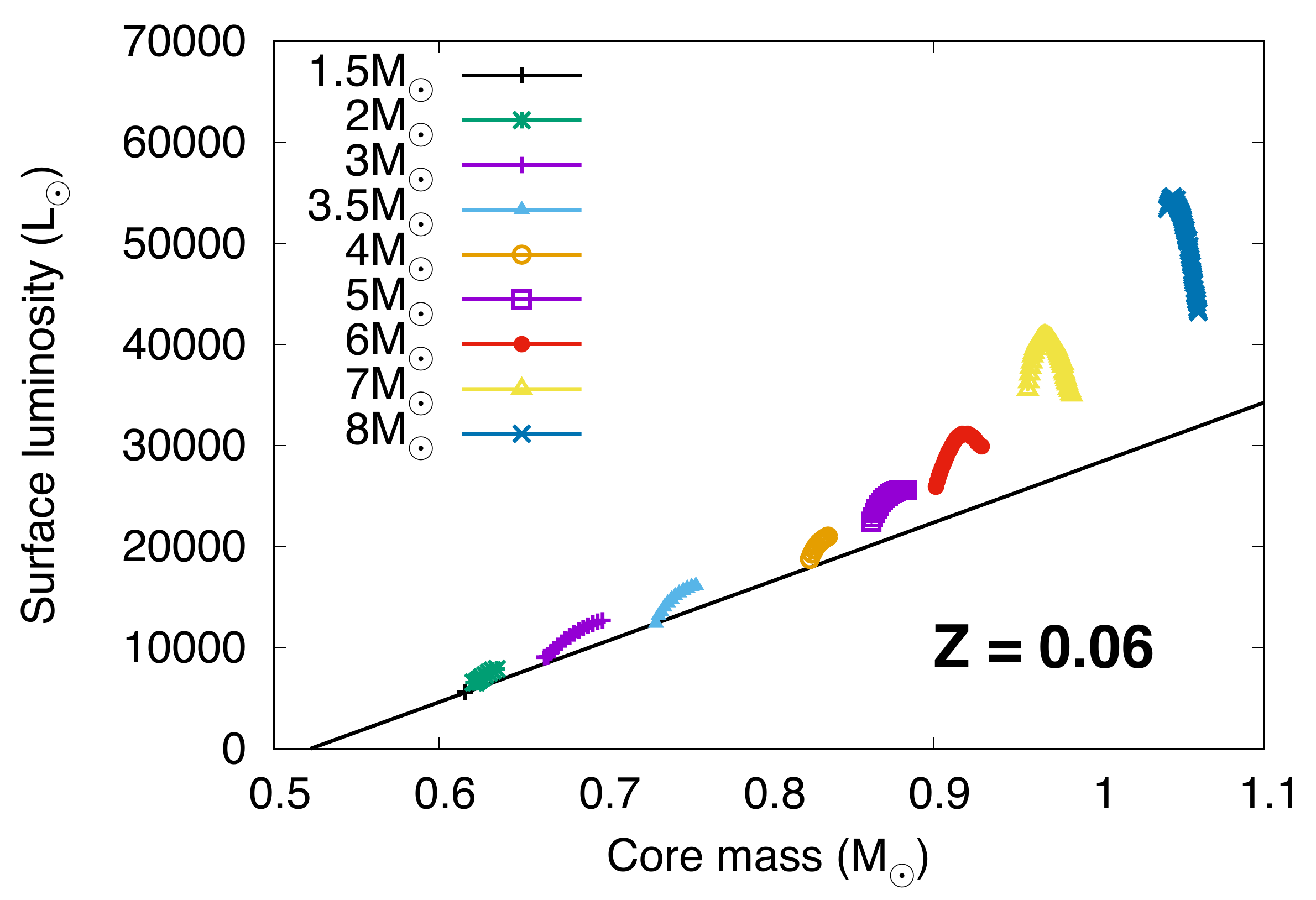}
 \caption{Stellar luminosity on the AGB as a function of the H-exhausted core mass for models of $Z=0.014$ (upper panel) and $Z=0.06$ (lower panel) models of 2$\Msun$ to 8$\Msun$. The 1$\Msun$ models are left out owing to the $Z=0.06$ case not making it to the TP-AGB phase. The solid black line is \citet{paczynski75}'s linear core-mass luminosity relationship. \label{fig:corelum}}
\end{center}
\end{figure}

The minimum mass for HBB is known to be metallicity dependent, with metal-poor AGB stars experiencing HBB at lower initial masses \citep[e.g.,][]{karakas10a,ventura09a,weiss09,colibri}. In models with $Z\ge 0.04$ we find that the minimum mass for HBB shifts from 4.5$\Msun$ for $Z=0.014$ to 5.5$\Msun$ for $Z= 0.04$ to 6.5$\Msun$ for $Z=0.09$. The only $Z=0.1$ models to experience HBB are the 7$\Msun$ and 7.5$\Msun$ models, which reach HBB temperatures on the early AGB before thermal pulses begin.  The main reason that metal-rich AGB models experience HBB at higher masses is owing to the structure at the base of the envelope, which is less dense than their metal-poor counterparts. Lower HBB temperatures has consequences for the maximum luminosity on the TP-AGB, as shown in Fig.~\ref{fig:corelum} for models of $Z=0.014$ \citep[data from][]{karakas14b} and $Z=0.06$, where we show the core-mass luminosity relationship for some of our models.  The linear relationship from \citet{paczynski75} is included for comparison and is
\begin{equation}
    L/\Lsun = 28320 + 59250 (M_{\rm c}/\Msun - 1.0),
\end{equation}
where $M_{\rm c}$ is the mass of the H-exhausted core. Interestingly the $Z=0.06$ models deviate less from the linear relationship compared to models of solar metallicity, and only for the higher masses $M\gtrsim 6\Msun$ with efficient HBB. From Fig~\ref{fig:corelum} we see that the $Z=0.014$ models with HBB exhibit higher peak luminosities on the AGB, as a result of their more efficient HBB. For example, the 8$\Msun$, $Z=0.014$ models reaches almost 64,000$\Lsun$, compared to 54,000$\Lsun$ for the $Z=0.06$ model of the same mass. The difference in luminosity almost directly correlates with the difference in peak temperatures reached at the base of the envelope. The 8$\Msun$, $Z=0.014$ model reaches over $\approx 10^{8}$~K, roughly 20\% higher than the peak temperature of the 8$\Msun$, $Z=0.06$ model, which is $\approx 0.82 \times 10^{8}$~K.

%new figure

%new sub-section
\subsection{Comparison to other metal-rich AGB models} \label{sec:compare}

We briefly compare the results of our TP-AGB models to other models published in the literature, starting with a comparison of AGB lifetimes of models up to $Z=0.06$ to those  published in \citet{marigo17} and \citet{ventura20}. The 3$\Msun$, $Z=0.04$ model in \citet{ventura20} has a total AGB lifetime of 14~Myr and a TP-AGB lifetime of 2.02~Myr. The \citet{ventura20} TP-AGB lifetime is considerably longer the 0.9~Myr lifetime found in our model of the same mass and $Z$.
However, the total AGB lifetimes are similar where $\tau_{\rm agb} = 13$~Myr in this paper. In contrast, the lowest mass models of 1$\Msun$ and the intermediate-mass models of $\approx 6\Msun$ have qualitatively similar lifetimes for both the entire AGB and TP-AGB phase compared to \citet{ventura20}.

\citet{marigo17} show the number of thermal pulses and lifetimes for a range of models of different metallicity, as a function of mass in their Fig.~2. Considering their most metal-rich models of [M/H] $=+0.7$, the 1.5$\Msun$ and 2$\Msun$ models have TP-AGB lifetimes of 0.2~Myr and 0.5~Myr, respectively, and experience roughly 5 and 12~TPs.  Our $Z=0.06$ AGB models have shorter TP-AGB lifetimes and fewer TPs. Our 2$\Msun$, $Z=0.06$ has an TP-AGB lifetime of only 0.15~Myr and 8 TPs, whereas the 1.5$\Msun$, $Z=0.06$ has only one TP and consequently a very short TP-AGB lifetime of $\approx 0.1$~Myr. These results are explained by our models experiencing stronger mass-loss early on in the AGB before the onset of thermal pulses, which erodes the envelope.

\citet{marigo17} provide the only AGB models up to [M/H] $\approx +0.7$ including maps of C-stars and O-rich stars, but they do not specifically comment on the occurrence of HBB in very metal-rich AGB stars. However \citet{marigo17} finds that C-stars are absent from their AGB models above [M/H] $\gtrsim 0.4$, consistent with our results. Similarly, none of the \citet{weiss09} models of $Z =0.04$ become C-rich either.  In contrast, \citet{ventura20} find their model of 2.5$\Msun$, $Z=0.04$ just becomes C-rich, with a final C/O $=1.08$. 

The core mass at the first thermal pulse, $M_{\rm c}(1)$, is an important indicator of the final stellar mass, especially for intermediate-mass stars that show little core growth on the AGB. \citet{weiss09} compare to the solar metallicity models (here $Z=0.02$) of \citet{karakas02} and found smaller cores for their low-mass AGB stars at the beginning of the TP-AGB; we find similar results here in comparison to their $Z=0.04$ models, but find very similar core masses for models around $4-6\Msun$.  The $Z=0.04$ models presented here also have larger core masses compared to the low-mass AGB models from \citet{ventura20}, but similar, although somewhat smaller, core masses for intermediate-masses (e.g., for 6$\Msun$ we find 0.895$\Msun$ in our model compared to 0.928$\Msun$ from Ventura et al and 0.893$\Msun$ from Weiss \& Ferguson).

The structure and nucleosynthesis of models with HBB is known to be notoriously dependent upon the temperature gradient in the convective envelope, and hence on the treatment of convection \citep[e.g.,][]{ventura05a}. We adopt the Mixing-length Theory (MLT) of convection which has been shown to produce lower peak temperatures and higher minimum masses for HBB than models that adopt the Full Spectrum of Turbulence \citep{ventura18}. However even stellar codes which adopt MLT show differences in their HBB models, e.g., the minimum mass for solar metallicity is 6$\Msun$ in \citet{weiss09}, around 5$\Msun$ for solar metallicity from \citet{colibri}, while HBB is not apparent in any of the solar metallicity models of \citet{straniero14}. 

\citet{ventura20} finds that the minimum mass for HBB is around 3.5-4$\Msun$ for $Z=0.04$, which is considerably lower than our 5.5$\Msun$. However, this difference is consistent with what we find at solar metallicity \citep{ventura18}. \citet{weiss09} find the minimum mass for HBB is 6$\Msun$ for $Z=0.04$, which is similar to our models, whereas \citet{colibri} find mild HBB in their 5$\Msun$, $Z= 0.04$ model, with some Li production, up to $\log \epsilon$(Li/H) $\approx 2$. 

\section{Discussion} \label{sec:discuss}

The reduction in the minimum mass for carbon burning from 8$\Msun$ to 7$\Msun$ in the $Z=0.1$ Monash models is an interesting result that has implications for a lowering of the core-collapse supernova threshold mass. The MESA models show that the minimum mass for C ignition is similarly reduced in metal-rich tracks to $7\Msun$ compared to $9\Msun$ at solar composition, with the most metal-rich model available ($Z=0.09$) showing a reduction to $6\Msun$. This drop in C ignition threshold to $6\Msun$ further applies to all metal-enhanced models $Z\ge 0.06$ when convective overshoot is used.

The minimum mass for C ignition is dependent upon a number of factors including the initial conditions \citep{bono00} and input physics including the temporal and spatial resolution \citep{siess06,doherty10}, and the inclusion of convective overshoot in the core during early evolutionary phases \citep{gilpons07}. The minimum mass for core collapse supernovae at very high metallicities is also dependent upon the choice of mass-loss, which may strip the envelope so fast that core-collapse is avoided \citep{ibeling13}. The models of \citet{mowlavi98a} suggest models of $Z=0.1$ can evolve toward core collapse, although the explosions would predominantly be of Types Ib/c, with no H present in the spectra.

Thermally-pulsing AGB models are known to be greatly affected by modelling uncertainties, both in the numerical details and in the input physics used in calculations \citep[e.g.,][]{ventura05a,ventura05b,stancliffe07a,karakas12,rosenfield14,pastorelli20,marigo20,TUMi,gilpons21}. The most significant for very metal-rich stars is the mass-loss rate, which is essentially unknown in evolved phases. Calculating convective temperature gradients and the borders between radiative and convective regions is an on-going problem that extends to these highest metallicities. 

In regards to the occurrence of the third dredge-up, the Monash models do not include any formal overshoot \citep[see discussion in][]{kamath12,lattanzio86}. Models without convective boundary mixing do not match observable features of C-stars, including the minimum mass for carbon star formation in Galactic and Magellanic Cloud populations \citep{marigo99}. However, this is contradicted by \citet{stancliffe05a} who match the carbon star luminosity function without convective overshoot. For this reason it is likely that our minimum initial stellar masses for TDU are likely to be overestimates. At solar metallicity the gap is around 0.5$\Msun$ between where we predict TDU to occur and the minimum initial mass for C-star formation \citep{karakas16}, but the difference could be as large as 1$\Msun$ if TDU occurs in 1$\Msun$ stars of solar metallicity as discussed below.

There has been progress in estimating the minimum initial stellar mass for the occurrence of third dredge-up in Galactic and Magallanic Cloud stellar populations. For Galactic carbon stars the initial mass range has been previously estimated to be between about 1.5--4$\Msun$ \citep{groen95,karakas14b,cristallo15,marigo20}. However, results using Gaia data shows that Galactic S-type stars, which show C/O $\approx 1$, are likely to have evolved from stars with initial masses closer to 1$\Msun$ \citep{shetye19}. These results are in contrast to \citet{marigo20} who found that the minimum initial mass for Milky Way carbon stars is around 1.65$\Msun$. For stars in the Magellanic Clouds, the minimum initial mass has been estimated to be as low as $1\Msun$ \citep{wood83} from stellar pulsations \citep{agbstars}, however the masses derived from pulsations are present-day masses, not initial. More recent theoretical work suggest that the minimum initial mass for C-stars in the LMC is instead around 1.7$\Msun$ \citep{pastorelli20}. Metal-poor post-AGB stars in the Magellanic Clouds are some of the most $s$-process rich objects known, where the stars rich in heavy elements are also carbon rich \citep{reyniers07a,vanaarle11,desmedt12,desmedt15}. The initial masses of the post-AGB stars are likely in the range 1--1.5$\Msun$, based on their luminosities, which are reasonably well known \citep{vanwinckel03,desmedt12}, also in tension with the results of \citet{pastorelli20}.  The discovery by \citet{kamath17} of a single luminous, low-metallicity post-AGB star in the Magellanic Clouds that appears to have failed TDU highlights that there is still much we do not understand about this phenomena.

Studies of Galactic and LMC evolved stellar populations highlight the issues surrounding the occurrence (or not) of third dredge-up, and the problem with trying to understand the behaviour as a function of metallicity. More metal-poor populations do become C-rich more easily and likely have more efficient TDU. If we turn this around to more metal-rich populations, a decreasing TDU efficiency is expected although the behaviour of the minimum initial stellar mass is unclear. In this study we find the minimum mass to be a strong function of the initial metallicity, to the point where TDU disappears from low-mass AGB populations by $Z=0.04$. 

We can experiment with the minimum mass for TDU by including convective overshoot in the Monash models in the same manner as described in \citet{karakas10b}. We ran  test calculations where we extend the base of the convective envelope following a thermal pulse by $N$ pressure-scale heights, where we set $N=1, 2, 3$ and 4. \citet{kamath12} found that values around $N=3-4$ were needed in order to match the luminosity of stars transitioning from oxygen rich to carbon rich in Magellanic Cloud clusters in stars around $1.5-2\Msun$. Focusing on the 3$\Msun$, $Z=0.04$ model as an example, the model with no formal overshoot has negligible dredge-up, where $\lambda < 0.1$. Setting $N = 1, 2, 3, 4$ produces $\lambda_{\rm max} = 0.2, 0.5, 0.8, 0.96$, respectively.  The model with $N=4$ becomes carbon rich, with a final C/O = 1.3, which highlights just how much TDU is needed in these high metallicity models in order to overcome the large amount of oxygen initially present.

Regarding AGB mass-loss, we adopt \citet{vw93} as in previous calculations \citep{karakas14b}. In lower metallicity models \citet{vw93} results in more thermal pulses and a longer TP-AGB lifetime than other AGB mass-loss formulations \citep[e.g.,][]{karakas14b}. In very-metal rich AGB models, \citet{vw93} leads to shorter or similar TP-AGB lifetimes for $Z=0.04$ compared to the \citet{bloecker95} mass-loss adopted by \citet{ventura20}. The \citet{bloecker95} mass-loss rate depends on a free parameter $\eta$, which \citet{ventura20} set  to $\eta = 0.02$ from calibration of their models to AGB stars in the Magellanic Clouds \citep{ventura01}. It is not clear how applicable such formulae or calibrations are to such metal-rich AGB stars. The same is true of course for the \citet{vw93} mass-loss rate we adopt.

We also find shorter TP-AGB lifetimes than \citet{marigo17}, where they adopt a staged prescription for mass-loss, depending upon the stage of AGB evolution. Early on in the AGB the mass-loss formulation of \citet{schroder05} is used, later when the star enters the dust-driven regime the mass-loss rate depends on the radius and mass, and is calibrated to Galactic long-period variables \citep{girardi10}. The superwind phase is then calculated in the same manner as \citet{vw93}. 

The main concern regarding AGB mass-loss are our results for $Z=0.1$, which have no to very few thermal pulses. One test is to turn off mass loss entirely, which provides a limit on behaviour of thermal pulses and HBB \citep[see e.g.,][]{karakas02}.  We test three models: the 2$\Msun$, 4$\Msun$ and 6.5$\Msun$ $Z=0.1$ models as examples. From Fig.~\ref{fig:agb} we see these models do not enter the TP-AGB phase.  The 2$\Msun$ model exits the AGB as a He-star and experiences no oscillations of the He-shell.  A closer examination of the He-shell luminosity of the 4$\Msun$ shows a few weak He-shell oscillations, which do not exceed $\log L/\Lsun \approx 3.7$, noting that the radiated luminosity at this stage is almost $\log L/\Lsun \approx 4.4$. In the case of the 6.5$\Msun$ model, He-shell flashes do occur with peak intensities of $\log L/\Lsun \approx 4.5$; these are below the radiated luminosity of $\log L/\Lsun \approx 4.6$. 

\begin{figure}
\begin{center}
\includegraphics[width=0.95\columnwidth]{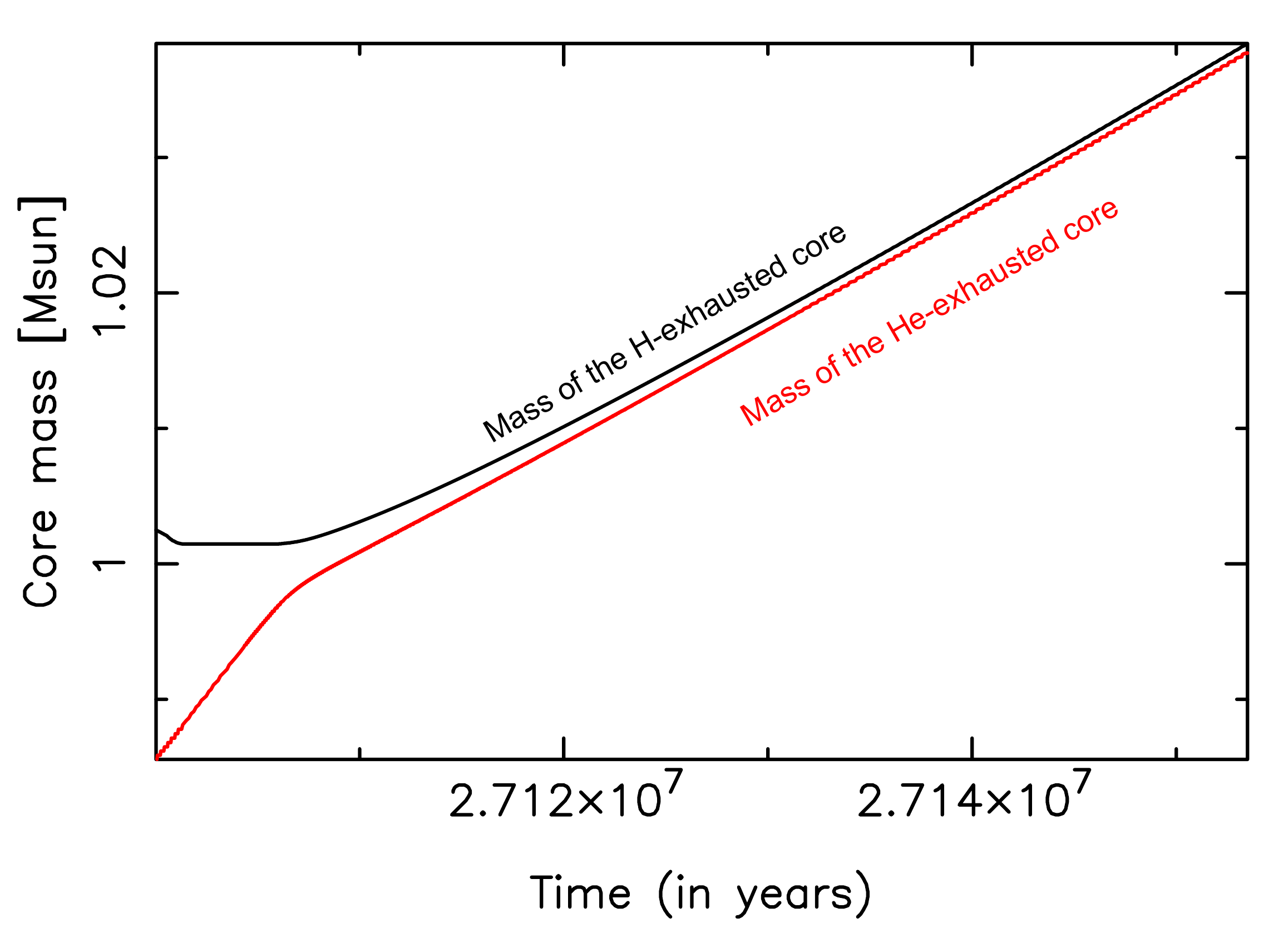}
\includegraphics[width=0.95\columnwidth]{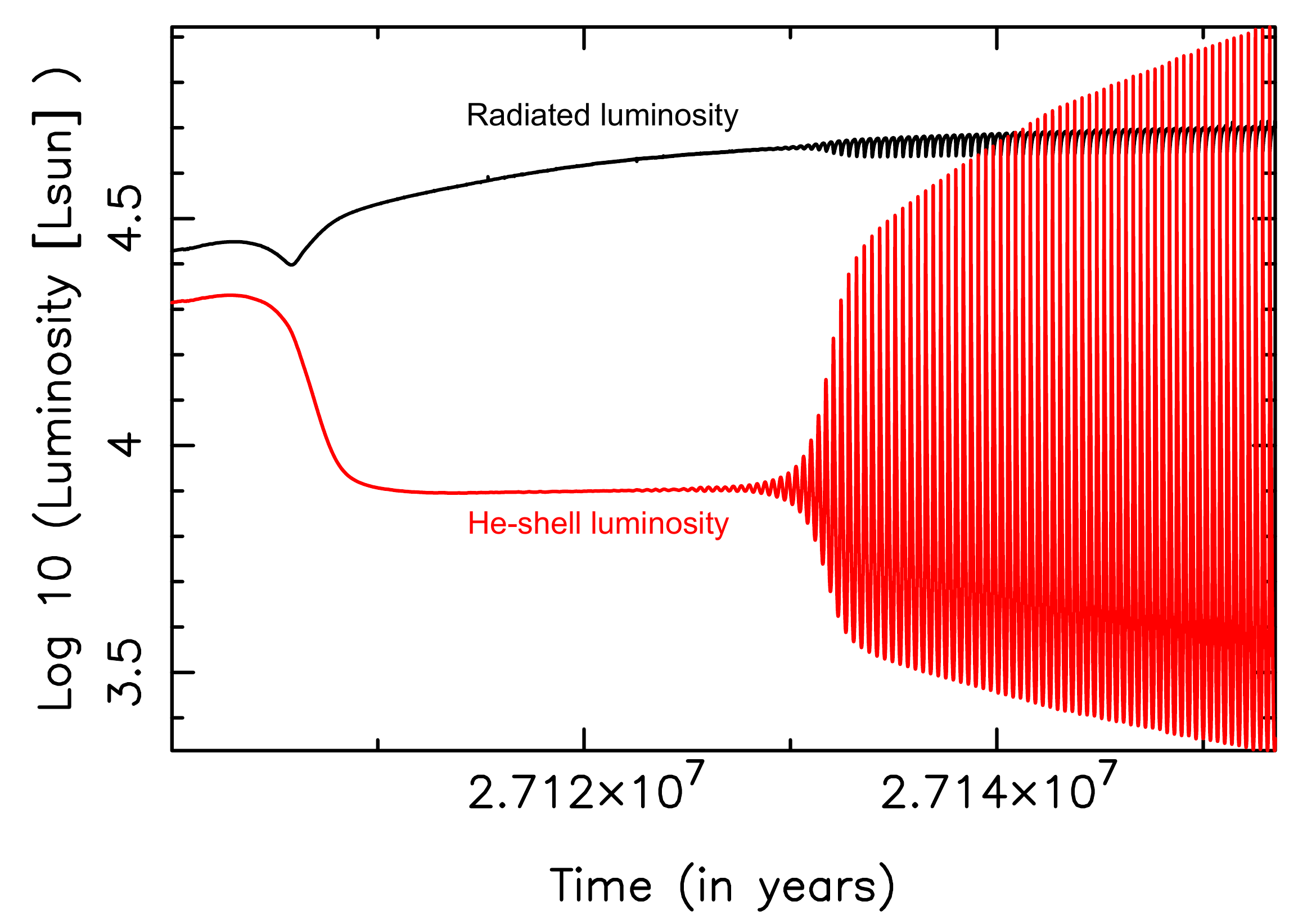}
 \caption{AGB evolution of the 6.5$\Msun$, $Z=0.1$ model with no mass-loss. The upper panel shows the evolution of the core during the TP-AGB phase, while the lower panel shows the radiated and He-shell luminosities. \label{fig:m6p5z10}}
\end{center}
\end{figure}

We evolve the models without mass-loss up to a maximum of 100,000 time-steps, noting that the models reach the early AGB in $\approx 1,000$. The 2$\Msun$ model enters the AGB with a H-exhausted core mass of 0.76$\Msun$ and we calculate 10 thermal pulses in total.
For the 4$\Msun$ model we calculate 44 thermal pulses, where the peak He-shell luminosity is $2.8\times 10^{5}\Lsun$ and still increasing. The maximum radiated luminosity during the interpulse is a fairly constant $\log L/\Lsun = 4.5$. No TDU or HBB is evident in either of these models.

In Fig~\ref{fig:m6p5z10} we show the evolution of the 6.5$\Msun$, $Z=0.1$ model with no mass-loss during the TP-AGB phase. We evolve through 37 thermal pulses, with Fig.~\ref{fig:m6p5z10} showing that the He-shell luminosity increases slowly with time, and only exceeds the radiated luminosity by the 28th thermal pulse. This model has HBB, with a peak temperature of about 75~MK but there is no evidence for TDU in the evolution of the core shown in Fig~\ref{fig:m6p5z10}. Interestingly, the maximum temperatures during thermal pulses is only $\approx 250$~MK in both the 4$\Msun$ and 6.5$\Msun$. For comparison, in solar metallicity models we find typically 350~MK for core masses $\approx 1\Msun$.

These experiments show that the very metal-rich models that exit the AGB as He-stars may experience thermal pulses with a different choice of mass-loss. This is entirely reasonable, given uncertainties about surrounding mass-loss in evolved stars \citep[e.g.][]{hoefner18}. These experiments also show that if the intermediate-mass models of $Z=0.1$ enter the TP-AGB they are likely to have HBB, albeit at lower temperatures compared to their more metal-poor counterparts. We rule out the occurrence of the TDU: Even after reaching an asymptotic behaviour these models experience thermal pulses that are just too weak to drive TDU mixing.

\section{Conclusions} \label{sec:conclusion}

In this paper we have presented new evolutionary tracks of very-metal rich AGB stars using two distinct stellar evolution codes. The Monash models cover a large range in mass, from 1 to 8$\Msun$, for metallicities between $Z=0.04$ to $Z=0.1$, evolved from the main sequence to near the end of the thermally-pulsing AGB phase. We have also included supplementary MESA models of intermediate-mass between $Z=0.06$ to $Z=0.09$, evolved to the beginning of the TP-AGB.  These are the first AGB models of $Z=0.1$ ($Z=0.09$) published in the literature. Our results are consistent with earlier $Z=0.1$ models by \citet{mowlavi98a,mowlavi98b} which show that the evolution of these most metal-rich stars is different to models of lower metallicity, and their behaviour cannot be easily extrapolated from models of lower metallicity. A good example of this is the minimum mass for C ignition, which shifts down from 8$\Msun$ to 7$\Msun$ in the Monash models of $Z=0.1$ and in MESA models with $Z=0.06-0.08$. We see an even deeper downward shift to lower masses in the most metal-rich MESA models and all MESA models with convective overshoot. 

Not all evolutionary features are strongly dependent on metallicity.
We find that the depth of second dredge-up in intermediate-mass stars is notably metallicity invariant. From Figure~\ref{fig:depth} we find SDU to be similar in the Monash models of solar metallicity all the way to our highest metallicity of $Z=0.1$. Even first dredge-up is reasonably metallicity invariant at the lower and upper mass end, although there are differences between about 2--4$\Msun$ when comparing models of solar metallicity, $Z=0.04$ and $Z=0.1$. The consequences of these results will be explored in Cinquegrana \& Karakas (2021) when we present surface abundances and stellar yields for very metal-rich evolved stellar populations.

In this paper we consider the metal-rich limit of AGB behaviour. We found that thermal pulses can occur in Monash models of up to $Z=0.1$ but that mass-loss on the AGB severely limits the number of thermal pulses in many models of such high metallicity. Third dredge-up, important for carbon and heavy-element production, ceases in models above $Z=0.08$, although HBB can in principle occur even in models of $Z=0.1$, depending upon the mass-loss rate on the AGB. This result has important implications for the chemical evolution of galaxies. Without a primary source of carbon the yields of such stars will be governed by secondary products mixed to the surface by first and second dredge-up, and secondary nitrogen made by HBB for models over about 6$\Msun$ and is discussed in detail in Cinquegrana \& Karakas (2021).

In this paper we considered the evolution of single stars. Very metal-rich stars are more extended in radius than their lower metallicity counterparts owing to their high opacities, so may interact with their binary companions in more interesting ways than their lower metallicity counterparts. Finally, given the limit of TDU at these high metallicities, nucleosynthesis during binary evolution (e.g., novae or Type Ia explosions) may be the only way for these stars to produce fresh metals and to have a significant impact on the chemical evolution of their host galaxies. 

\section*{Acknowledgements}
We thank the Referee for their detailed comments which have helped to improve the paper, and GC is grateful for discussions with Carolyn Doherty on modelling super-AGB stars.
This research was supported by the Australian Research Council Centre of Excellence for All Sky Astrophysics in 3 Dimensions (ASTRO 3D), through project number CE170100013. 
M.J.\ was supported the Research School of Astronomy and Astrophysics at the Australian National University and funding from Australian Research Council grant No.\ DP150100250 as well as the Lasker Fellowship at the Space Telescope Science Institute. MJ wishes to thank Rob Farmer for useful discussions and Pablo Marchant for use of his Kippenhahn diagram Python package, available at 
\verb|https://doi.org/10.5281/zenodo.2602097|.

%%%%%%%%%%%%%%%%%%%%%%%%%%%%%%%%%%%%%%%%%%%%%%%%%%
\section*{Data Availability}

In the case of the Monash models, the data underlying this article will be shared on request to the corresponding author. 
The MESA tracks and accompanying control files will be made publicly available at a Zenodo repository link after processing of this manuscript.



\bibliographystyle{mnras}
\bibliography{mnemonic,library,Joyce_bib_2021}

%%%%%%%%%%%%%%%%%%%%%%%%%%%%%%%%%%%%%%%%%%%%%%%%%%

%%%%%%%%%%%%%%%%% APPENDICES %%%%%%%%%%%%%%%%%%%%%

\appendix

\section{MESA Thresholds}

Here we provide a pictorial chart showing the details of the MESA models calculated for this paper. We include the initial mass, initial metallicity $Z_{\rm in}$, initial helium mass fraction $Y_{\rm in}$, whether carbon burning was detected or not, the location of the burning,  whether there was a pocket of higher convective burning closer to the core, whether the burning was vigorous, and the maximum age of the model (in Myr). We refer to the text in Section~\ref{sec:mesa} for details. Solar composition models are included for comparison to the metal-rich intermediate-mass models (e.g., see Fig~\ref{fig:hrd-m6}).

\begin{figure*}
\begin{center}
\includegraphics{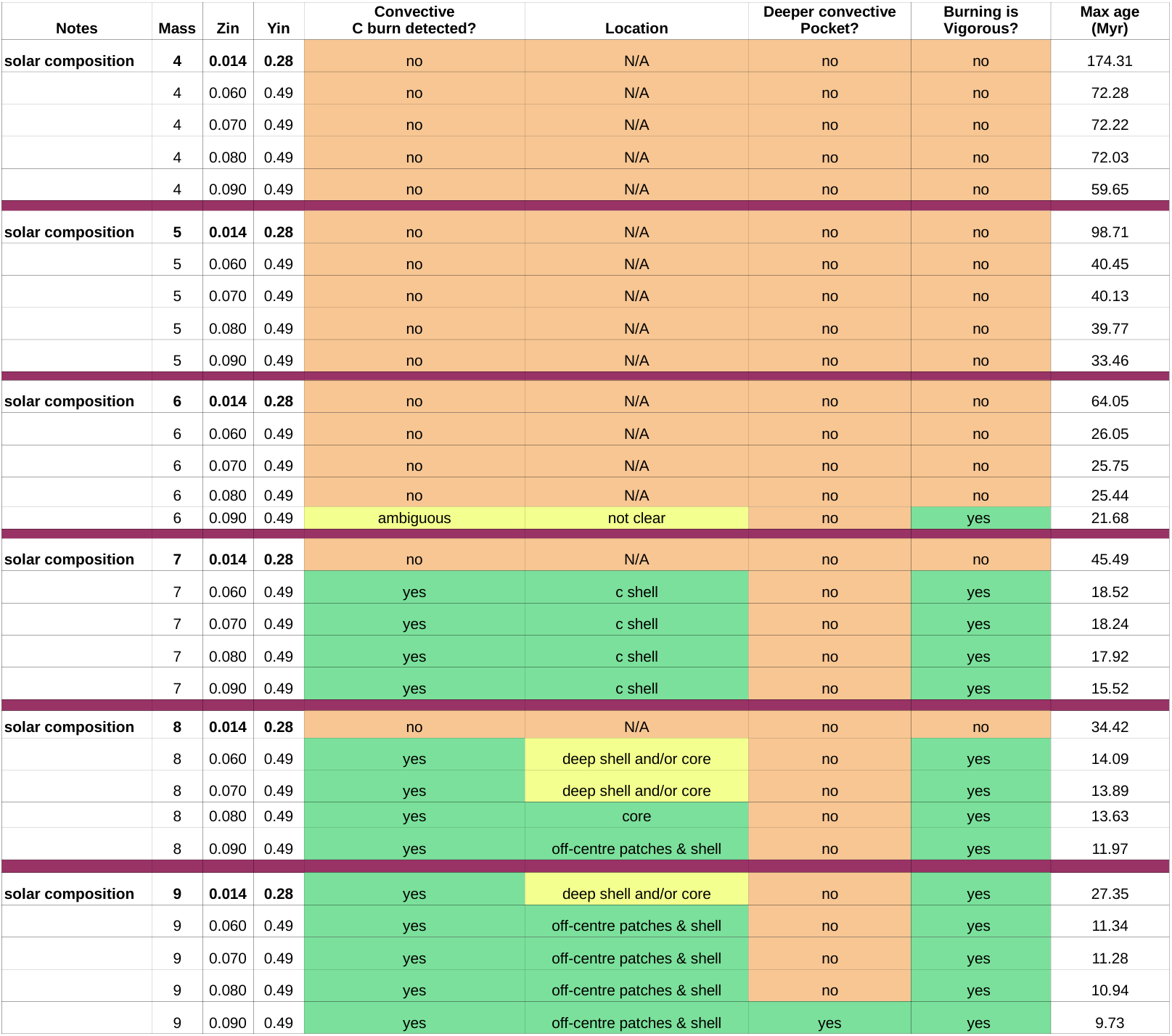}
 \caption{Chart depicting carbon ignition among MESA models as a function of mass and metallicity ($Z$). For each mass, a model computed with solar abundance is included for reference. The detection of carbon burning is performed by inspection of Kippenhahn diagrams and $T_{\text{max}}-\rho$ diagnostic plots. In some cases, the precise location and definition of the core are unclear; ambiguity is indicated with yellow boxes. 
 \label{fig:cMESA}}
\end{center}
\end{figure*}

\section{AGB model data}

Tables of thermally-pulsing AGB model data for metallicities of $Z=0.05$ to $Z=0.1$.

\begin{table*}
\begin{center}
\caption{Stellar models calculated for $Z=0.05$, $Z=0.06$ and $Z=0.07$.}
\label{tab:modelsz06-z07}
\begin{tabular}{lcccccccccc} \hline \hline
Mass & SDU & \#TP & $\lambda_{\rm max}$ & $M_{\rm c}(1)$ & 
$T_{\rm bce}^{\rm max}$ & $\log L_{\rm agb}^{\rm max}$ & 
$\log L_{\rm He-shell}^{\rm max}$ & $\tau_{\rm stellar}$ & 
$\tau_{\rm rgb}$ & $\tau_{\rm agb}$ \\
($\Msun$) &  &   &  & ($\Msun$) & (MK) & ($\Lsun$) & ($\Lsun$) & (Myr) & (Myr) & (Myr) \\ \hline
\multicolumn{11}{c}{$Z=0.05$, $Y=0.354$ models.} \\ \hline
1.0 & No & 0 & 0.00 & -- & -- & 3.556 & -- & 14110 & 3406 & 20 \\
1.5 & No & 4 & 0.00 & 0.603 & 2.5 & 3.763 & 4.88 & 3222 & 756.2 & 20 \\ 
2.0 & No &  14 & 0.00 & 0.595 & 3.3 & 3.898 & 5.55 & 1340 & 148.7  & 19 \\
2.5 & No &  21 & 0.00 & 0.604 & 4.0 & 4.010 & 5.95 & 787.2 & 37.2 & 18 \\
3.0 & No &  22 & 0.00 & 0.636 & 5.1 & 4.106 & 6.26 & 450.6 & 16.8 & 12 \\
3.5 & No &  17 & 0.00 & 0.700 & 6.5 & 4.192 & 6.43 & 276.0 & 9.6 & 7.2 \\ 
4.0 & No &  12 & 0.05 & 0.787 & 10 & 4.288 & 6.46 & 182.8 & 6.0 & 4.5  \\
4.5 & Yes & 12 & 0.32 & 0.840 & 18 & 4.359 & 6.62 & 129.8 & 4.0 & 2.9 \\
5.0 & Yes & 19 & 0.62 & 0.857 & 28 & 4.403 & 7.01 & 96.5 & 2.8 & 2.1 \\
5.5 & Yes & 26 & 0.74 & 0.874 & 47 & 4.448 & 7.26 & 75.4 & 2.0 & 1.5 \\
5.75 & Yes & 29 & 0.74 & 0.887 & 56 & 4.477 & 7.26 & 66.9 & 1.7 & 1.3 \\
6.0 & Yes & 34 & 0.76 & 0.897 & 62 & 4.505 & 7.10 & 53.8 & 1.4 & 1.2 \\
6.5 & Yes & 39 & 0.80 & 0.915 & 68 & 4.561 & 7.36 & 49.4 & 1.0 & 0.91 \\
7.0 & Yes & 44 & 0.80 & 0.953 & 73 & 4.616 & 7.39 & 40.8 & 0.75 & 0.70 \\
7.5 & Yes & 51 & 0.79 & 0.991 & 79 & 4.675 & 7.35 & 35.2 & 0.58 & 0.53 \\
8.0 & Yes$^{\rm a}$ & 59 & 0.80 & 1.034 & 86 & 4.742 & 7.38 & 30.0 & 0.42 & 0.44 \\ \hline
\multicolumn{11}{c}{$Z=0.06$, $Y=0.375$ models.} \\ \hline
1.0 & No & 0 & 0.00 & -- & -- & 3.550 & -- & 12900 & 3431 & 21 \\
1.5 & No & 1 & 0.00 & 0.616 & 2.5 & 3.747 & 4.38 & 2986 & 665 & 17 \\
2.0 & No & 8 & 0.00 & 0.621 & 3.2 & 3.898 & 4.97 & 1265 & 128 & 19 \\
2.5 & No & 15 & 0.00 & 0.631 & 3.9 & 4.009 & 5.55 & 721.7 & 33.1 & 15 \\
3.0 & No & 15 & 0.00 & 0.664 & 4.8 & 4.104 & 5.84 & 408.4 & 15.5 & 10\\
3.5 & No & 12 & 0.00 & 0.732 & 6.0 & 4.208 & 6.06 & 252.3 & 8.87 & 6.1 \\
4.0 & Yes & 9 & 0.00 & 0.825 & 7.8 & 4.322 & 6.10 & 171.0 & 5.49 & 3.5 \\
4.5 & Yes & 14 & 0.00 & 0.843 & 16 & 4.370 & 6.24 & 118.1 & 3.73 & 2.7\\
5.0 & Yes & 18 & 0.04 & 0.862 & 23 & 4.409 & 6.36 & 88.5 & 2.58 & 1.9 \\
5.5 & Yes & 24 & 0.39 & 0.882 & 36 & 4.449 & 6.53 & 68.4 & 1.81 & 1.5 \\
6.0 & Yes & 35 & 0.56 & 0.901 & 55 & 4.494 & 6.79 & 53.8 & 1.37 & 1.1 \\
6.5 & Yes & 41 & 0.55 & 0.926 & 65 & 4.555 & 6.75 & 44.8 & 0.94 & 0.91 \\
7.0 & Yes & 48 & 0.62 & 0.957 & 67 & 4.611 & 6.93 & 37.6 & 0.69 & 0.67 \\
7.5 & Yes & 52 & 0.58 & 0.994 & 76 & 4.668 & 6.76 & 31.4 & 0.50 & 0.52 \\
8.0 & Yes$^{\rm a}$ & 66 & 0.64 & 1.041 & 82 & 4.735 & 6.79 & 27.2 & 0.39 & 0.42 \\ \hline
\multicolumn{11}{c}{$Z=0.07$, $Y=0.396$ models.} \\ \hline
1.0 & No & 0 & 0.00 & -- & -- & 3.47 & -- & 11788 & 3377 & 433.776\\
1.5 & No & 0 & 0.00 & -- & -- & 3.71 & -- & 2692 & 566.1 & 14.66 \\
2.0 & No & 2 & 0.00 & 0.651 & 3.2 & 3.925 & 4.53 & 1142 & 102.2 & 16.06 \\ 
2.5 & No & 10 & 0.00 & 0.662 & 4.0 & 4.045 & 5.14 & 636.5 & 29.85 & 12.89 \\
3.0 & No & 12 & 0.00 & 0.697 & 5.1 & 4.145 & 5.57 & 359.7 & 14.13 & 8.943\\
3.5 & No & 9 & 0.00 & 0.772 & 6.3 & 4.260 & 5.80 & 222.5 & 7.865 & 5.280 \\
4.0 & Yes & 9 & 0.00 & 0.836 & 8.1 & 4.350 & 5.86 & 145.1 & 4.716 & 3.181\\
4.5 & Yes & 16 & 0.00 & 0.852 & 18 & 4.394 & 6.07 & 105.7 & 3.368 & 2.409 \\
5.0 & Yes & 22 & 0.00 & 0.874 & 27 & 4.436 & 6.11 & 76.65 & 2.082 & 1.671\\
5.5 & Yes & 30 & 0.11 & 0.892 & 41 & 4.475 & 6.23 & 61.28 & 1.525 & 1.289 \\
6.0 & Yes & 40 & 0.26 & 0.914 & 59 & 4.527 & 6.34 & 48.75 & 1.059 & 1.036\\
6.5 & Yes & 49 & 0.29 & 0.940 & 66 & 4.582 & 6.36 & 40.00 & 0.761 & 0.801 \\
7.0 & Yes & 56 & 0.31 & 0.967 & 71 & 4.634 & 6.36 & 33.16 & 0.546 & 0.665\\
7.5 & Yes & 61 & 0.29 & 1.003 & 75 & 4.684 & 6.30 & 28.35 & 0.401 & 0.508\\
8.0 & Yes$^{\rm a}$ & 73 & 0.34 & 1.049 & 82 & 4.753 & 6.32 & 23.86 & 0.297 & 0.418\\
\hline \hline
\end{tabular}
\medskip\\
\noindent (a) Model ignited carbon on early AGB, with core temperatures nearing or exceeding 900~MK.
\end{center}
\end{table*}

\begin{table*}
\begin{center}
\caption{Stellar models calculated for $Z=0.08$, $Z=0.09$ and $Z=0.1$.}
\label{tab:modelsz08-z10}
\begin{tabular}{lcccccccccc} \hline \hline
Mass & SDU & \#TP & $\lambda_{\rm max}$ & $M_{\rm c}(1)$ & 
$T_{\rm bce}^{\rm max}$ & $\log L_{\rm agb}^{\rm max}$ & 
$\log L_{\rm He-shell}^{\rm max}$ & $\tau_{\rm stellar}$ & 
$\tau_{\rm rgb}$ & $\tau_{\rm agb}$ \\
($\Msun$) &  &   &  & ($\Msun$) & (MK) & ($\Lsun$) & ($\Lsun$) & (Myr) & (Myr) & (Myr) \\ \hline
\multicolumn{11}{c}{$Z=0.08$, $Y=0.420$ models.} \\ \hline
1.0 & No & 0 & 0.00 & -- & -- & 3.51 & -- & 10012 & 2833 & 415.7\\
1.5 & No & 0 & 0.00 & -- & -- & 3.72 & -- & 2332 & 459.2 & 12.44 \\
2.0 & No & 0 & 0.00 & -- & -- & 3.89 & -- & 1063 & 86.29 & 11.70 \\
2.5 & No & 1 & 0.00 & 0.694 & 3.8 & 4.038 & 4.32 & 558.4 & 26.80 & 9.881 \\
3.0 & No & 2 & 0.00 & 0.746 & 4.0 & 4.167 & 4.70 & 318.7 & 12.55 & 6.073\\
3.5 & No & 3 & 0.00 & 0.819 & 5.2 & 4.296 & 5.090 & 196.1 & 7.094 & 4.178\\
4.0 & Yes & 6 & 0.00 & 0.845 & 6.6 & 4.358 & 5.34 & 127.5 & 4.253 & 2.845\\
4.5 & Yes & 11 & 0.00 & 0.863 & 13 & 4.403 & 5.64 & 92.58 & 2.997 & 2.075\\
5.0 & Yes & 16 & 0.00 & 0.887 & 20 & 4.446 & 5.76 & 68.0 & 2.0 & 1.48 \\
5.0 & Yes & 16 & 0.00 & 0.887 & 20 & 4.447 & 5.76 & 67.99 & 1.842 & 1.484\\
5.5 & Yes & 26 & 0.00 & 0.905 & 31 & 4.481 & 5.91 & 53.80 & 1.356 & 1.181\\
6.0 & Yes & 33 & 0.00 & 0.933 & 48 & 4.526 & 5.91 & 42.25 & 0.908 & 0.883\\
6.5 & Yes & 45 & 0.00 & 0.962 & 63 & 4.587 & 5.96 & 35.33 & 0.651 & 0.676\\
7.0 & Yes & 53 & 0.09 & 0.988 & 69 & 4.635 & 6.06 & 29.19 & 0.472 & 0.574\\
7.5 & Yes & 61 & 0.15 & 1.024 & 75 & 4.692 & 6.02 & 24.82 & 0.348 & 0.474\\
8.0 & Yes$^{\rm a}$ & 46 & 0.00 & 1.094 & 84 & 4.788 & 5.61 & 21.37 & 0.263 & 0.370\\
 \hline
\multicolumn{11}{c}{$Z=0.09$, $Y=0.438$ models.} \\ \hline
1.0 & No & 0 & 0.00 & -- & -- & 3.49 & -- & 8762 & 2494 & 381.8\\
1.5 & No & 0 & 0.00 & -- & -- & 3.74 & -- & 2073 & 386.3 & 11.33\\
2.0 & No & 0 & 0.00 & -- & -- & 3.90 & -- & 912.4 & 70.52 & 9.938\\
2.5 & No & 0 & 0.00 & -- & -- & 3.66 & -- & 481.8 & 23.44 & 7.665\\
3.0 & No & 0 & 0.00 & -- & -- & 3.84 & -- & 273.1 & 11.15 & 5.152\\
3.5 & No & 1 & 0.00 & 0.834 & 0.002 & 4.311 & 4.44 & 168.7 & 6.445 & 3.691\\
4.0 & Yes & 1 & 0.00 & 0.858 & 4.9 & 4.368 & 4.70 & 111.7 & 3.805 & 2.341\\
4.5 & Yes & 6 & 0.00 & 0.878 & 7.3 & 4.416 & 5.11 & 80.85 & 2.575 & 1.711\\
5.0 & Yes & 11 & 0.00 & 0.901 & 15 & 4.460 & 5.31 & 59.11 & 1.614 & 1.306\\
5.5 & Yes & 18 & 0.00 & 0.924 & 24 & 4.498 & 5.46 & 47.19 & 1.183 & 0.950\\
6.0 & Yes & 25 & 0.00 & 0.953 & 39 & 4.539 & 5.62 & 37.55 & 0.801 & 0.743\\
6.5 & Yes & 34 & 0.00 & 0.985 & 58 & 4.589 & 5.56 & 30.99 & 0.574 & 0.587\\
7.0 & Yes & 30 & 0.00 & 1.017 & 68 & 4.648 & 5.36 & 25.26 & 0.404 & 0.489\\
7.5 & Yes$^{\rm a}$ & 22 & 0.00 & 1.058 & 76 & 4.72 & 5.170 & 21.73 & 0.299 & 0.410\\
8.0 & Yes$^{\rm a}$ & 0 & 0.00 & -- & -- & 4.733 & -- & 18.68 & 0.225 & 0.304\\ \hline
\multicolumn{11}{c}{$Z=0.1$, $Y=0.459$ models.} \\ \hline
1.0 & No & 0 & 0.00 & -- & -- & 3.790 & -- & 7033 & 2050 & 15 \\
1.5 & No & 0 & 0.00 & -- & -- & 3.830 & -- & 1779 & 308.4 & 7.9 \\
2.0 & No & 0 & 0.00 & -- & -- & 3.990 & -- & 790.4 & 59.5 & 7.0 \\
2.5 & No & 0 & 0.00 & -- & 3.4 & 4.110 & -- & 410.1 & 20.3 & 6.0 \\
3.0 & No & 0 & 0.00 & -- & 3.6 & 4.240 & -- & 234.0 & 10.0 & 3.8 \\
3.5 & Yes & 0 & 0.00 & -- & 4.9 & 4.317 & -- & 146.0 & 5.8 & 2.8 \\
4.0 & Yes & 0 & 0.00 & -- & 6.0 & 4.375 & -- & 98.1 & 3.8 & 2.0 \\
4.5 & Yes & 0 & 0.00 & 0.890 & 7.0 & 4.430 & 3.96 & 70.3 & 2.4 & 1.4 \\
5.0 & Yes & 5 & 0.00 & 0.917 & 14 & 4.476 & 4.68 & 52.5 & 1.6 & 1.1 \\
5.5 & Yes & 7 & 0.00 & 0.945 & 20 & 4.516 & 4.73 & 40.9 & 1.1 & 0.81 \\
6.0 & Yes & 7 & 0.00 & 0.981 & 32 & 4.563 & 4.72 & 32.5 & 0.80 & 0.62 \\
6.5 & Yes & 4 & 0.00 & -- & 45 & 4.610 & -- & 26.6 & 0.57 & 0.48 \\
7.0 & Yes$^{\rm a}$ & 0 & 0.00 & -- & 68 & 4.674 & -- & 22.8 & 0.42 & 0.37 \\
7.5 & Yes$^{\rm a}$ & 0 & 0.00 & -- & -- & -- & -- & 19.6 & 0.30 & 0.28 \\
8.0 & Yes$^{\rm a}$ & 0 & 0.00 & -- & -- & -- & -- & 17.2 & 0.23 & 0.22 \\
\hline \hline
\end{tabular}
\medskip\\
\noindent (a) Model ignited carbon on early AGB, with core temperatures nearing or exceeding 900~MK.
\end{center}
\end{table*}

%6.5 Z =0.1, max He-shell luminosity =4.52 < radiated.

\section{HR Diagrams}

\begin{figure}
\begin{center}
\includegraphics[width=0.95\columnwidth]{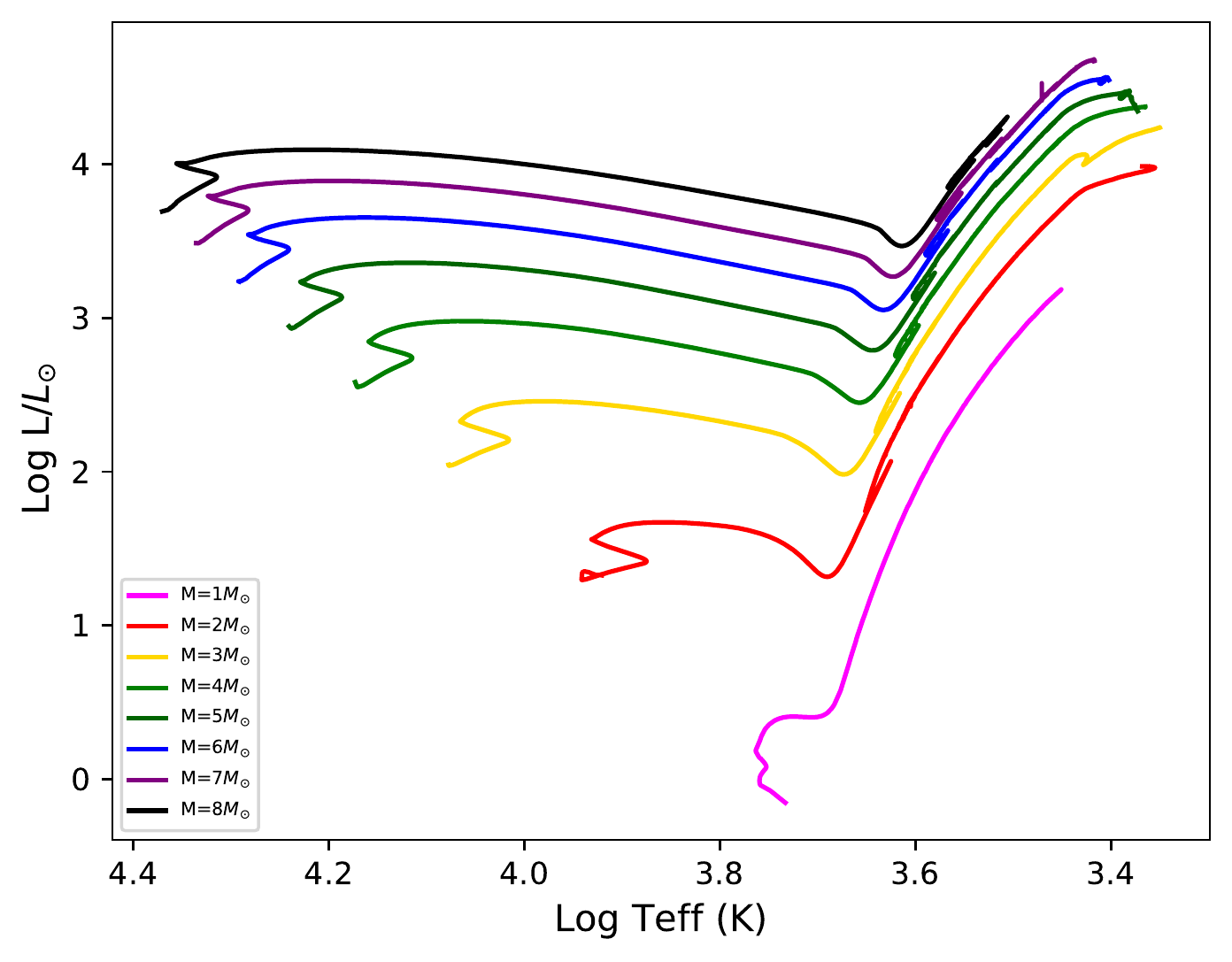}
\includegraphics[width=0.95\columnwidth]{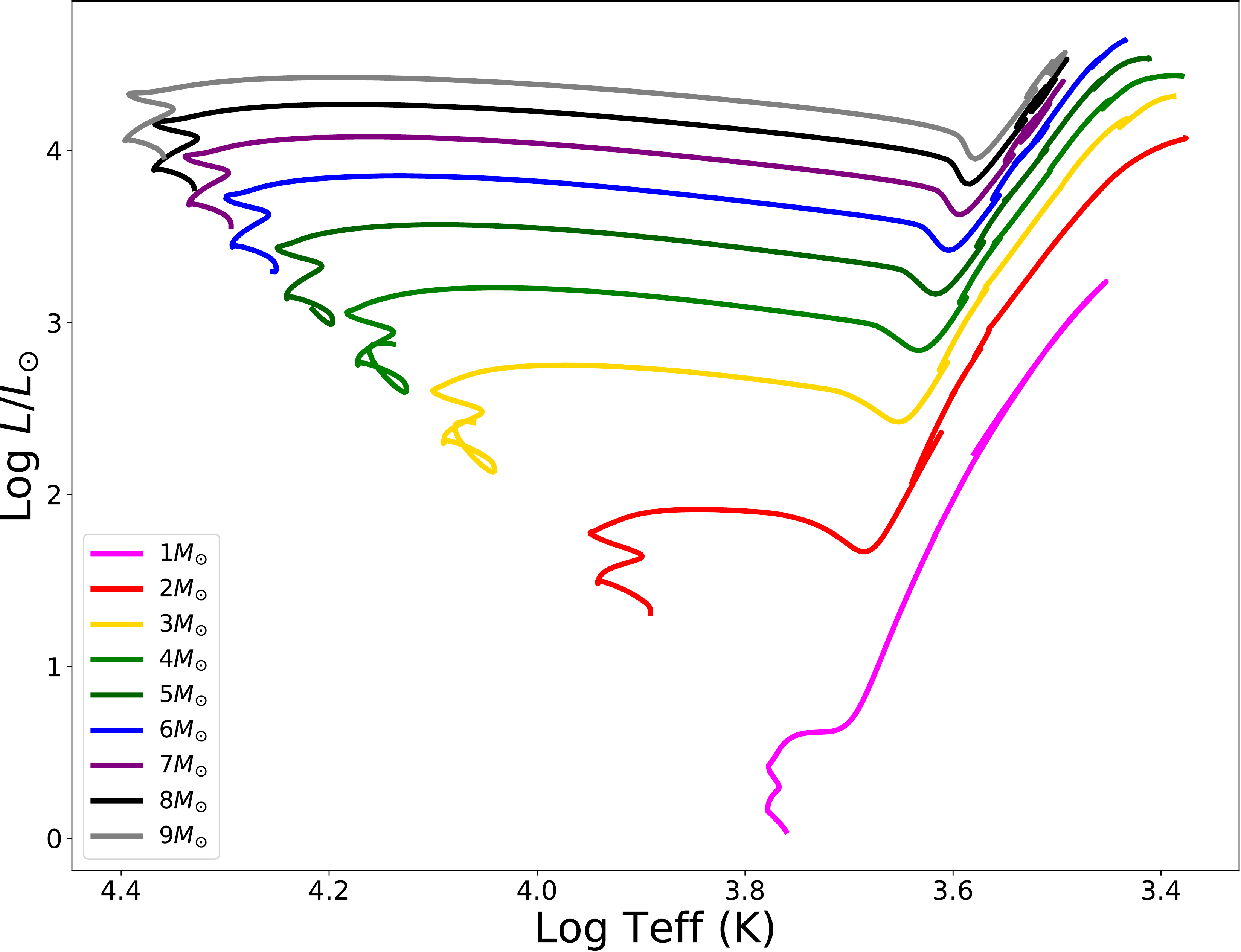}
 \caption{ Evolutionary tracks for a selection of the Monash $Z=0.1$ models (upper panel) and MESA $Z=0.09$ models (lower panel). Equivalent masses are presented in the same colour in both panels.
 \label{fig:hrd-z10}}
\end{center}
\end{figure}

\begin{figure}
\begin{center}
\includegraphics[width=0.95\columnwidth]{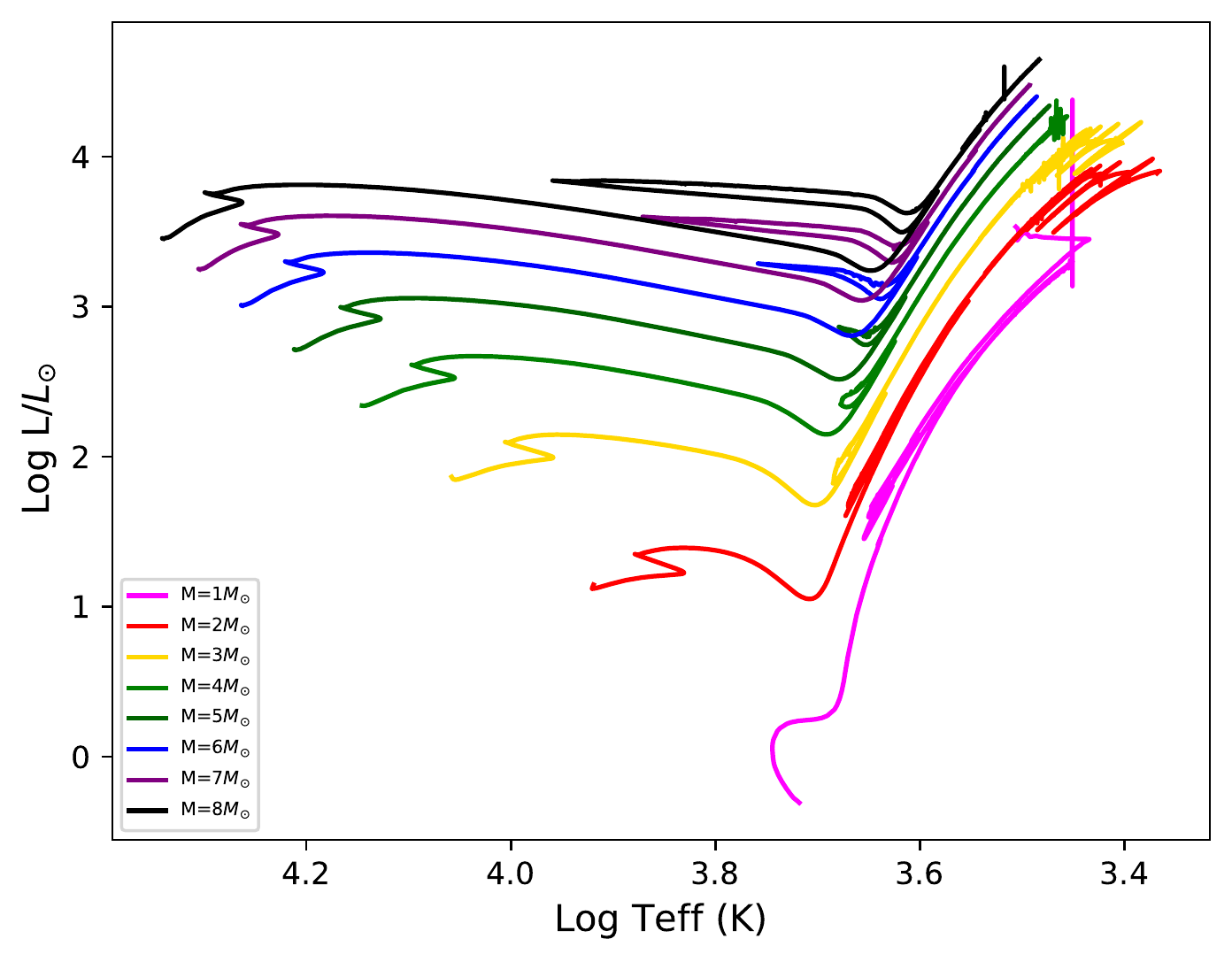}
 \caption{Evolutionary tracks for a selection of the $Z=0.04$ Monash models. \label{fig:hrd-z04}}
\end{center}
\end{figure}

\begin{figure}
\begin{center}
\includegraphics[width=0.95\columnwidth]{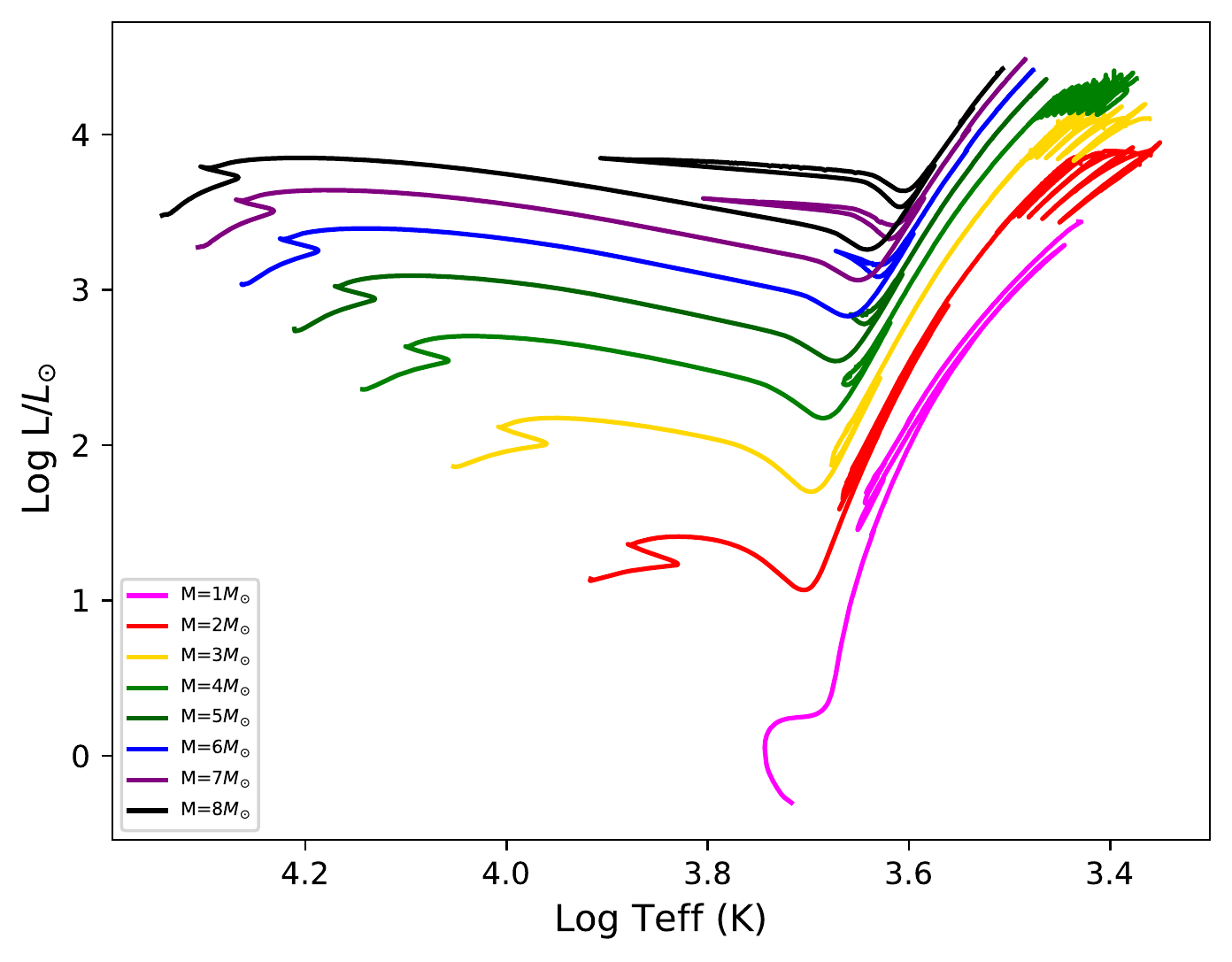}
 \caption{Same as Fig.~\ref{fig:hrd-z04} for $Z=0.05$. \label{fig:hrd-z05}}
\end{center}
\end{figure}

\begin{figure}
\begin{center}
\includegraphics[width=0.95\columnwidth]{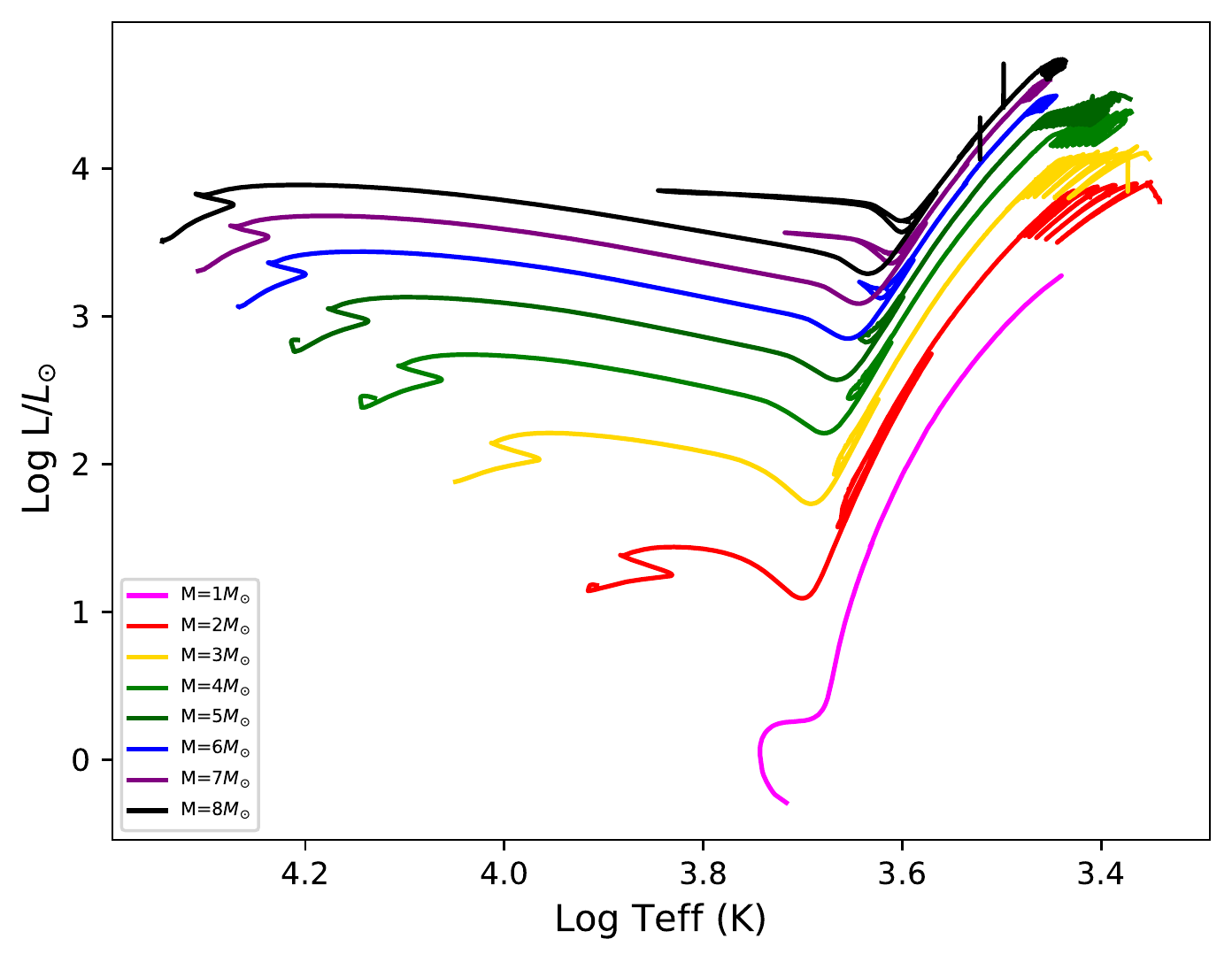}
 \caption{Same as Fig.~\ref{fig:hrd-z04} for $Z=0.06$. \label{fig:hrd-z06}}
\end{center}
\end{figure}

In Figures~\ref{fig:hrd-z04}, ~\ref{fig:hrd-z05} and~\ref{fig:hrd-z06} we show HR diagrams for a selection of the Monash models of $Z=0.04$, $Z=0.05$ and $Z=0.06$.

%%%%%%%%%%%%%%%%%%%%%%%%%%%%%%%%%%%%%%%%%%%%%%%%%%

% Don't change these lines
\bsp	% typesetting comment
\label{lastpage}
\end{document}